\ifx\undefined\psfig\else\endinput\fi

\edef\psfigRestoreAt{\catcode`@=\number\catcode`@\relax}
\catcode`\@=11\relax
\newwrite\@unused
\def\ps@typeout#1{{\let\protect\string\immediate\write\@unused{#1}}}
\ps@typeout{psfig/tex 1.8}

\def\figurepath{./}
\def\psfigurepath#1{\edef\figurepath{#1}}

\def\@nnil{\@nil}
\def\@empty{}
\def\@psdonoop#1\@@#2#3{}
\def\@psdo#1:=#2\do#3{\edef\@psdotmp{#2}\ifx\@psdotmp\@empty \else
    \expandafter\@psdoloop#2,\@nil,\@nil\@@#1{#3}\fi}
\def\@psdoloop#1,#2,#3\@@#4#5{\def#4{#1}\ifx #4\@nnil \else
       #5\def#4{#2}\ifx #4\@nnil \else#5\@ipsdoloop #3\@@#4{#5}\fi\fi}
\def\@ipsdoloop#1,#2\@@#3#4{\def#3{#1}\ifx #3\@nnil
       \let\@nextwhile=\@psdonoop \else
      #4\relax\let\@nextwhile=\@ipsdoloop\fi\@nextwhile#2\@@#3{#4}}
\def\@tpsdo#1:=#2\do#3{\xdef\@psdotmp{#2}\ifx\@psdotmp\@empty \else
    \@tpsdoloop#2\@nil\@nil\@@#1{#3}\fi}
\def\@tpsdoloop#1#2\@@#3#4{\def#3{#1}\ifx #3\@nnil
       \let\@nextwhile=\@psdonoop \else
      #4\relax\let\@nextwhile=\@tpsdoloop\fi\@nextwhile#2\@@#3{#4}}
\ifx\undefined\fbox
\newdimen\fboxrule
\newdimen\fboxsep
\newdimen\ps@tempdima
\newbox\ps@tempboxa
\long\def\fbox#1{\leavevmode\setbox\ps@tempboxa\hbox{#1}\ps@tempdima\fboxrule
    \advance\ps@tempdima \fboxsep \advance\ps@tempdima \dp\ps@tempboxa
   \hbox{\lower \ps@tempdima\hbox
  {\vbox{\hrule height \fboxrule
          \hbox{\vrule width \fboxrule \hskip\fboxsep
          \vbox{\vskip\fboxsep \box\ps@tempboxa\vskip\fboxsep}\hskip
                 \fboxsep\vrule width \fboxrule}
                 \hrule height \fboxrule}}}}
\fi
\newread\ps@stream
\newif\ifnot@eof       
\newif\if@noisy        
\newif\if@atend        
\newif\if@psfile       
{\catcode`\%=12\global\gdef\epsf@start{
\def\epsf@PS{PS}
\def\epsf@getbb#1{%
\openin\ps@stream=#1
\ifeof\ps@stream\ps@typeout{Error, File #1 not found}\else
   {\not@eoftrue \chardef\other=12
    \def\do##1{\catcode`##1=\other}\dospecials \catcode`\ =10
    \loop
       \if@psfile
	  \read\ps@stream to \epsf@fileline
       \else{
	  \obeyspaces
          \read\ps@stream to \epsf@tmp\global\let\epsf@fileline\epsf@tmp}
       \fi
       \ifeof\ps@stream\not@eoffalse\else
       \if@psfile\else
       \expandafter\epsf@test\epsf@fileline:. \\%
       \fi
          \expandafter\epsf@aux\epsf@fileline:. \\%
       \fi
   \ifnot@eof\repeat
   }\closein\ps@stream\fi}%
\long\def\epsf@test#1#2#3:#4\\{\def\epsf@testit{#1#2}
			\ifx\epsf@testit\epsf@start\else
\ps@typeout{Warning! File does not start with `\epsf@start'.  It may not be a
PostScript file.}
			\fi
			\@psfiletrue} 
{\catcode`\%=12\global\let\epsf@percent=
\long\def\epsf@aux#1#2:#3\\{\ifx#1\epsf@percent
   \def\epsf@testit{#2}\ifx\epsf@testit\epsf@bblit
	\@atendfalse
        \epsf@atend #3 . \\%
	\if@atend
	   \if@verbose{
		\ps@typeout{psfig: found `(atend)'; continuing search}
	   }\fi
        \else
        \epsf@grab #3 . . . \\%
        \not@eoffalse
        \global\no@bbfalse
        \fi
   \fi\fi}%
\def\epsf@grab #1 #2 #3 #4 #5\\{%
   \global\def\epsf@llx{#1}\ifx\epsf@llx\empty
      \epsf@grab #2 #3 #4 #5 .\\\else
   \global\def\epsf@lly{#2}%
   \global\def\epsf@urx{#3}\global\def\epsf@ury{#4}\fi}%
\def\epsf@atendlit{(atend)}
\def\epsf@atend #1 #2 #3\\{%
   \def\epsf@tmp{#1}\ifx\epsf@tmp\empty
      \epsf@atend #2 #3 .\\\else
   \ifx\epsf@tmp\epsf@atendlit\@atendtrue\fi\fi}

\chardef\letter = 11
\chardef\other = 12

\newif \ifdebug 
\newif\ifc@mpute 
\c@mputetrue 

\let\then = \relax
\def\r@dian{pt }
\let\r@dians = \r@dian
\let\dimensionless@nit = \r@dian
\let\dimensionless@nits = \dimensionless@nit
\def\internal@nit{sp }
\let\internal@nits = \internal@nit
\newif\ifstillc@nverging
\def \Mess@ge #1{\ifdebug \then \message {#1} \fi}

{ 
	\catcode `\@ = \letter
	\gdef \nodimen {\expandafter \n@dimen \the \dimen}
	\gdef \term #1 #2 #3%
	       {\edef \t@ {\the #1}
		\edef \t@@ {\expandafter \n@dimen \the #2\r@dian}%
		\t@rm {\t@} {\t@@} {#3}%
	       }
	\gdef \t@rm #1 #2 #3%
	       {{%
		\count 0 = 0
		\dimen 0 = 1 \dimensionless@nit
		\dimen 2 = #2\relax
		\Mess@ge {Calculating term #1 of \nodimen 2}%
		\loop
		\ifnum	\count 0 < #1
		\then	\advance \count 0 by 1
			\Mess@ge {Iteration \the \count 0 \space}%
			\Multiply \dimen 0 by {\dimen 2}%
			\Mess@ge {After multiplication, term = \nodimen 0}%
			\Divide \dimen 0 by {\count 0}%
			\Mess@ge {After division, term = \nodimen 0}%
		\repeat
		\Mess@ge {Final value for term #1 of
				\nodimen 2 \space is \nodimen 0}%
		\xdef \Term {#3 = \nodimen 0 \r@dians}%
		\aftergroup \Term
	       }}
	\catcode `\p = \other
	\catcode `\t = \other
	\gdef \n@dimen #1pt{#1} 
}

\def \Divide #1by #2{\divide #1 by #2} 

\def \Multiply #1by #2
       {{
	\count 0 = #1\relax
	\count 2 = #2\relax
	\count 4 = 65536
	\Mess@ge {Before scaling, count 0 = \the \count 0 \space and
			count 2 = \the \count 2}%
	\ifnum	\count 0 > 32767 
	\then	\divide \count 0 by 4
		\divide \count 4 by 4
	\else	\ifnum	\count 0 < -32767
		\then	\divide \count 0 by 4
			\divide \count 4 by 4
		\else
		\fi
	\fi
	\ifnum	\count 2 > 32767 
	\then	\divide \count 2 by 4
		\divide \count 4 by 4
	\else	\ifnum	\count 2 < -32767
		\then	\divide \count 2 by 4
			\divide \count 4 by 4
		\else
		\fi
	\fi
	\multiply \count 0 by \count 2
	\divide \count 0 by \count 4
	\xdef \product {#1 = \the \count 0 \internal@nits}%
	\aftergroup \product
       }}

\def\r@duce{\ifdim\dimen0 > 90\r@dian \then   
		\multiply\dimen0 by -1
		\advance\dimen0 by 180\r@dian
		\r@duce
	    \else \ifdim\dimen0 < -90\r@dian \then  
		\advance\dimen0 by 360\r@dian
		\r@duce
		\fi
	    \fi}

\def\Sine#1%
       {{%
	\dimen 0 = #1 \r@dian
	\r@duce
	\ifdim\dimen0 = -90\r@dian \then
	   \dimen4 = -1\r@dian
	   \c@mputefalse
	\fi
	\ifdim\dimen0 = 90\r@dian \then
	   \dimen4 = 1\r@dian
	   \c@mputefalse
	\fi
	\ifdim\dimen0 = 0\r@dian \then
	   \dimen4 = 0\r@dian
	   \c@mputefalse
	\fi
	\ifc@mpute \then
		\divide\dimen0 by 180
		\dimen0=3.141592654\dimen0
		\dimen 2 = 3.1415926535897963\r@dian 
		\divide\dimen 2 by 2 
		\Mess@ge {Sin: calculating Sin of \nodimen 0}%
		\count 0 = 1 
		\dimen 2 = 1 \r@dian 
		\dimen 4 = 0 \r@dian 
		\loop
			\ifnum	\dimen 2 = 0 
			\then	\stillc@nvergingfalse
			\else	\stillc@nvergingtrue
			\fi
			\ifstillc@nverging 
			\then	\term {\count 0} {\dimen 0} {\dimen 2}%
				\advance \count 0 by 2
				\count 2 = \count 0
				\divide \count 2 by 2
				\ifodd	\count 2 
				\then	\advance \dimen 4 by \dimen 2
				\else	\advance \dimen 4 by -\dimen 2
				\fi
		\repeat
	\fi
			\xdef \sine {\nodimen 4}%
       }}

\def\Cosine#1{\ifx\sine\UnDefined\edef\Savesine{\relax}\else
		             \edef\Savesine{\sine}\fi
	{\dimen0=#1\r@dian\advance\dimen0 by 90\r@dian
	 \Sine{\nodimen 0}
	 \xdef\cosine{\sine}
	 \xdef\sine{\Savesine}}}

\def\psdraft{
	\def\@psdraft{0}
}
\def\psfull{
	\def\@psdraft{100}
}

\psfull

\newif\if@scalefirst
\def\psscalefirst{\@scalefirsttrue}
\def\psrotatefirst{\@scalefirstfalse}
\psrotatefirst

\newif\if@draftbox
\def\psnodraftbox{
	\@draftboxfalse
}
\def\psdraftbox{
	\@draftboxtrue
}
\@draftboxtrue

\newif\if@prologfile
\newif\if@postlogfile
\def\pssilent{
	\@noisyfalse
}
\def\psnoisy{
	\@noisytrue
}
\psnoisy
\newif\if@bbllx
\newif\if@bblly
\newif\if@bburx
\newif\if@bbury
\newif\if@height
\newif\if@width
\newif\if@rheight
\newif\if@rwidth
\newif\if@angle
\newif\if@clip
\newif\if@verbose
\def\@p@@sclip#1{\@cliptrue}

\newif\if@decmpr

\def\@p@@sfigure#1{\def\@p@sfile{null}\def\@p@sbbfile{null}
	        \openin1=#1.bb
		\ifeof1\closein1
	        	\openin1=\figurepath#1.bb
			\ifeof1\closein1
			        \openin1=#1
				\ifeof1\closein1%
				       \openin1=\figurepath#1
					\ifeof1
					   \ps@typeout{Error, File #1 not found}
						\if@bbllx\if@bblly
				   		\if@bburx\if@bbury
			      				\def\@p@sfile{#1}%
			      				\def\@p@sbbfile{#1}%
							\@decmprfalse
				  	   	\fi\fi\fi\fi
					\else\closein1
				    		\def\@p@sfile{\figurepath#1}%
				    		\def\@p@sbbfile{\figurepath#1}%
						\@decmprfalse
	                       		\fi%
			 	\else\closein1%
					\def\@p@sfile{#1}
					\def\@p@sbbfile{#1}
					\@decmprfalse
			 	\fi
			\else
				\def\@p@sfile{\figurepath#1}
				\def\@p@sbbfile{\figurepath#1.bb}
				\@decmprtrue
			\fi
		\else
			\def\@p@sfile{#1}
			\def\@p@sbbfile{#1.bb}
			\@decmprtrue
		\fi}

\def\@p@@sfile#1{\@p@@sfigure{#1}}

\def\@p@@sbbllx#1{
		\@bbllxtrue
		\dimen100=#1
		\edef\@p@sbbllx{\number\dimen100}
}
\def\@p@@sbblly#1{
		\@bbllytrue
		\dimen100=#1
		\edef\@p@sbblly{\number\dimen100}
}
\def\@p@@sbburx#1{
		\@bburxtrue
		\dimen100=#1
		\edef\@p@sbburx{\number\dimen100}
}
\def\@p@@sbbury#1{
		\@bburytrue
		\dimen100=#1
		\edef\@p@sbbury{\number\dimen100}
}
\def\@p@@sheight#1{
		\@heighttrue
		\dimen100=#1
   		\edef\@p@sheight{\number\dimen100}
}
\def\@p@@swidth#1{
		\@widthtrue
		\dimen100=#1
		\edef\@p@swidth{\number\dimen100}
}
\def\@p@@srheight#1{
		\@rheighttrue
		\dimen100=#1
		\edef\@p@srheight{\number\dimen100}
}
\def\@p@@srwidth#1{
		\@rwidthtrue
		\dimen100=#1
		\edef\@p@srwidth{\number\dimen100}
}
\def\@p@@sangle#1{
		\@angletrue
		\edef\@p@sangle{#1} 
}
\def\@p@@ssilent#1{
		\@verbosefalse
}
\def\@p@@sprolog#1{\@prologfiletrue\def\@prologfileval{#1}}
\def\@p@@spostlog#1{\@postlogfiletrue\def\@postlogfileval{#1}}
\def\@cs@name#1{\csname #1\endcsname}
\def\@setparms#1=#2,{\@cs@name{@p@@s#1}{#2}}
\def\ps@init@parms{
		\@bbllxfalse \@bbllyfalse
		\@bburxfalse \@bburyfalse
		\@heightfalse \@widthfalse
		\@rheightfalse \@rwidthfalse
		\def\@p@sbbllx{}\def\@p@sbblly{}
		\def\@p@sbburx{}\def\@p@sbbury{}
		\def\@p@sheight{}\def\@p@swidth{}
		\def\@p@srheight{}\def\@p@srwidth{}
		\def\@p@sangle{0}
		\def\@p@sfile{} \def\@p@sbbfile{}
		\def\@p@scost{10}
		\def\@sc{}
		\@prologfilefalse
		\@postlogfilefalse
		\@clipfalse
		\if@noisy
			\@verbosetrue
		\else
			\@verbosefalse
		\fi
}
\def\parse@ps@parms#1{
	 	\@psdo\@psfiga:=#1\do
		   {\expandafter\@setparms\@psfiga,}}
\newif\ifno@bb
\def\bb@missing{
	\if@verbose{
		\ps@typeout{psfig: searching \@p@sbbfile \space  for bounding box}
	}\fi
	\no@bbtrue
	\epsf@getbb{\@p@sbbfile}
        \ifno@bb \else \bb@cull\epsf@llx\epsf@lly\epsf@urx\epsf@ury\fi
}
\def\bb@cull#1#2#3#4{
	\dimen100=#1 bp\edef\@p@sbbllx{\number\dimen100}
	\dimen100=#2 bp\edef\@p@sbblly{\number\dimen100}
	\dimen100=#3 bp\edef\@p@sbburx{\number\dimen100}
	\dimen100=#4 bp\edef\@p@sbbury{\number\dimen100}
	\no@bbfalse
}
\newdimen\p@intvaluex
\newdimen\p@intvaluey
\def\rotate@#1#2{{\dimen0=#1 sp\dimen1=#2 sp
		  \global\p@intvaluex=\cosine\dimen0
		  \dimen3=\sine\dimen1
		  \global\advance\p@intvaluex by -\dimen3
		  \global\p@intvaluey=\sine\dimen0
		  \dimen3=\cosine\dimen1
		  \global\advance\p@intvaluey by \dimen3
		  }}
\def\compute@bb{
		\no@bbfalse
		\if@bbllx \else \no@bbtrue \fi
		\if@bblly \else \no@bbtrue \fi
		\if@bburx \else \no@bbtrue \fi
		\if@bbury \else \no@bbtrue \fi
		\ifno@bb \bb@missing \fi
		\ifno@bb \ps@typeout{FATAL ERROR: no bb supplied or found}
			\no-bb-error
		\fi
		\count203=\@p@sbburx
		\count204=\@p@sbbury
		\advance\count203 by -\@p@sbbllx
		\advance\count204 by -\@p@sbblly
		\edef\ps@bbw{\number\count203}
		\edef\ps@bbh{\number\count204}
		\if@angle
			\Sine{\@p@sangle}\Cosine{\@p@sangle}
	        	{\dimen100=\maxdimen\xdef\r@p@sbbllx{\number\dimen100}
					    \xdef\r@p@sbblly{\number\dimen100}
			                    \xdef\r@p@sbburx{-\number\dimen100}
					    \xdef\r@p@sbbury{-\number\dimen100}}
                        \def\minmaxtest{
			   \ifnum\number\p@intvaluex<\r@p@sbbllx
			      \xdef\r@p@sbbllx{\number\p@intvaluex}\fi
			   \ifnum\number\p@intvaluex>\r@p@sbburx
			      \xdef\r@p@sbburx{\number\p@intvaluex}\fi
			   \ifnum\number\p@intvaluey<\r@p@sbblly
			      \xdef\r@p@sbblly{\number\p@intvaluey}\fi
			   \ifnum\number\p@intvaluey>\r@p@sbbury
			      \xdef\r@p@sbbury{\number\p@intvaluey}\fi
			   }
			\rotate@{\@p@sbbllx}{\@p@sbblly}
			\minmaxtest
			\rotate@{\@p@sbbllx}{\@p@sbbury}
			\minmaxtest
			\rotate@{\@p@sbburx}{\@p@sbblly}
			\minmaxtest
			\rotate@{\@p@sbburx}{\@p@sbbury}
			\minmaxtest
			\edef\@p@sbbllx{\r@p@sbbllx}\edef\@p@sbblly{\r@p@sbblly}
			\edef\@p@sbburx{\r@p@sbburx}\edef\@p@sbbury{\r@p@sbbury}
		\fi
		\count203=\@p@sbburx
		\count204=\@p@sbbury
		\advance\count203 by -\@p@sbbllx
		\advance\count204 by -\@p@sbblly
		\edef\@bbw{\number\count203}
		\edef\@bbh{\number\count204}
}
\def\in@hundreds#1#2#3{\count240=#2 \count241=#3
		     \count100=\count240	
		     \divide\count100 by \count241
		     \count101=\count100
		     \multiply\count101 by \count241
		     \advance\count240 by -\count101
		     \multiply\count240 by 10
		     \count101=\count240	
		     \divide\count101 by \count241
		     \count102=\count101
		     \multiply\count102 by \count241
		     \advance\count240 by -\count102
		     \multiply\count240 by 10
		     \count102=\count240	
		     \divide\count102 by \count241
		     \count200=#1\count205=0
		     \count201=\count200
			\multiply\count201 by \count100
		 	\advance\count205 by \count201
		     \count201=\count200
			\divide\count201 by 10
			\multiply\count201 by \count101
			\advance\count205 by \count201
		     \count201=\count200
			\divide\count201 by 100
			\multiply\count201 by \count102
			\advance\count205 by \count201
		     \edef\@result{\number\count205}
}
\def\compute@wfromh{
		\in@hundreds{\@p@sheight}{\@bbw}{\@bbh}
		\edef\@p@swidth{\@result}
}
\def\compute@hfromw{
	        \in@hundreds{\@p@swidth}{\@bbh}{\@bbw}
		\edef\@p@sheight{\@result}
}
\def\compute@handw{
		\if@height
			\if@width
			\else
				\compute@wfromh
			\fi
		\else
			\if@width
				\compute@hfromw
			\else
				\edef\@p@sheight{\@bbh}
				\edef\@p@swidth{\@bbw}
			\fi
		\fi
}
\def\compute@resv{
		\if@rheight \else \edef\@p@srheight{\@p@sheight} \fi
		\if@rwidth \else \edef\@p@srwidth{\@p@swidth} \fi
}
\def\compute@sizes{
	\compute@bb
	\if@scalefirst\if@angle
	\if@width
	   \in@hundreds{\@p@swidth}{\@bbw}{\ps@bbw}
	   \edef\@p@swidth{\@result}
	\fi
	\if@height
	   \in@hundreds{\@p@sheight}{\@bbh}{\ps@bbh}
	   \edef\@p@sheight{\@result}
	\fi
	\fi\fi
	\compute@handw
	\compute@resv}

\def\psfig#1{\vbox {
	\ps@init@parms
	\parse@ps@parms{#1}
	\compute@sizes
	\ifnum\@p@scost<\@psdraft{
		\special{ps::[begin] 	\@p@swidth \space \@p@sheight \space
				\@p@sbbllx \space \@p@sbblly \space
				\@p@sbburx \space \@p@sbbury \space
				startTexFig \space }
		\if@angle
			\special {ps:: \@p@sangle \space rotate \space}
		\fi
		\if@clip{
			\if@verbose{
				\ps@typeout{(clip)}
			}\fi
			\special{ps:: doclip \space }
		}\fi
		\if@prologfile
		    \special{ps: plotfile \@prologfileval \space } \fi
		\if@decmpr{
			\if@verbose{
				\ps@typeout{psfig: including \@p@sfile.Z \space }
			}\fi
			\special{ps: plotfile "`zcat \@p@sfile.Z" \space }
		}\else{
			\if@verbose{
				\ps@typeout{psfig: including \@p@sfile \space }
			}\fi
			\special{ps: plotfile \@p@sfile \space }
		}\fi
		\if@postlogfile
		    \special{ps: plotfile \@postlogfileval \space } \fi
		\special{ps::[end] endTexFig \space }
		\vbox to \@p@srheight true sp{
			\hbox to \@p@srwidth true sp{
				\hss
			}
		\vss
		}
	}\else{
		\if@draftbox{
			\hbox{\frame{\vbox to \@p@srheight true sp{
			\vss
			\hbox to \@p@srwidth true sp{ \hss \@p@sfile \hss }
			\vss
			}}}
		}\else{
			\vbox to \@p@srheight true sp{
			\vss
			\hbox to \@p@srwidth true sp{\hss}
			\vss
			}
		}\fi

	}\fi
}}
\psfigRestoreAt

\documentstyle[12pt]{article}
\sloppy
\pagestyle{plain}

\newtheorem{proposition}{Proposition}
\newtheorem{lemma}{Lemma}
\newcommand{\be}{\begin{equation}}
\newcommand{\ee}{\end{equation}}
\newcommand{\bd}{\begin{displaymath}}
\newcommand{\ed}{\end{displaymath}}
\newcommand{\lb}{\label}
\newcommand{\Bbb}{\rm}
\input psfig
\psfigurepath{.}
\typeout{Bilder sind aus \figurepath}

\begin{document}
\begin{titlepage}
\begin{flushright}
Z\"urich University Preprint \\
ZU-TH 7/95\\
\end{flushright}
\begin{center}

\vskip 1.5cm

{\LARGE Numerical Solutions of the}\\

\bigskip

{\LARGE Einstein-Yang-Mills System with}\\

\medskip

{\LARGE Cosmological Constant}
\vskip 1.5cm
{\large P\'{a}l G\'{e}za Moln\'{a}r} \footnote{Electronic address:
molnar@hirschen.physik.unizh.ch}
\vskip 0.4cm
{\it Institute for Theoretical Physics, University of Z\"urich, }\\
{\it Winter\-thur\-er\-stras\-se\nobreak\ 190, 8057 Z\"urich, Switzerland}

\vskip 0.4cm
\end{center}
{\bf Abstract:} Numerical evidence for a cosmological version of the
Bartnik-McKinnon family of particle-like solutions of the
Einstein-Yang-Mills system is presented. Our solutions are also
static, but space has the topology of a three-sphere.\\
By adjusting the cosmological constant we found numerically a
spherically symmetric solution which can be regarded as an
excitation of the unique SO$(4)$-invariant solution. We expect
that for each node number there exists such a solution without
a cosmological horizon.

\begin{center}
\vskip 0.4cm
{\large Diploma thesis}\\ {\large written under the supervision of}\\
{\large Prof. Dr. N. Straumann}\\
{\large Winter 1994/95}
\end{center}
\end{titlepage}

\newpage
\thispagestyle{empty}
\mbox{}

\newpage
\thispagestyle{empty}
\protect
\section*{}

\vskip 4cm

\begin{flushright}
{\Large\bf Az ap\'{a}mnak}
\end{flushright}

\newpage
\thispagestyle{empty}
\tableofcontents
\newpage
%
%
\section{Introduction}
\pagenumbering{arabic}
In recent years, relativists have been researching on what
happens when gravity is coupled to nonlinear f\/ield theories,
as for example the Yang-Mills (YM) f\/ields. In the beginning,
there was doubt about whether there were any solutions,
since neither the vacuum-Einstein equations \cite{Li}, nor the
pure YM equations have nontrivial, static, globally regular
solutions \cite{De,SC}.

Thus, it was even more surprising, as in 1988 Bartnik and
McKinnon (BK) discovered soliton solutions to the SU$(2)$
Einstein-Yang-Mills (EYM) system numerically \cite{BK}. These
solutions provided for the f\/irst time solid evidence that
the EYM system has particle-like solutions. Later, other
authors \cite{VG,PB,KM} found the colored black hole solutions to
the same model, which are a counterexample to the no-hair
conjecture; i.e., the well-known uniqueness theorem for stationary
black holes of the Einstein-Maxwell system has no natural
generalisation on the non-Abelian case. Soon, the existence of
both types of solution was established rigorously \cite{SWY1,SW,SWY2,Breit},
and their instability was proven \cite{SZ1,SZ2,SZ3,Z}. Because of
this instability, it was expected that it would also apply to
arbitrary gauge groups. Proof was f\/irst successfully provided
for the EYM solitons \cite{OS1} and shortly afterwards for black
holes as well \cite{OS2}.

There were also attempts to f\/ind corresponding solutions for
other related f\/ield theories. The Einstein-Skyrme system for
example has black hole solutions with hair that are at least
li\-ne\-arly stable \cite{DHS1,DHS2,DHS3,HS}. Several authors looked
at other models, notably the SU$(2)$ Einstein-Yang-Mills-Higgs
(EYMH) theory with a Higgs triplet
\cite{LNW,BFM,AB}, as well as the EYM-dilaton theory \cite{DG}.
Recently, the instability of the gravitating, regular sphaleron
solutions to the SU$(2)$ EYMH system with a SU$(2)$ Higgs doublet,
that were known numerically for some time \cite{GMN},
could be shown \cite{BBMSV}.

In November, 1993, Ding and Hosoya (DH) presented a spherically
symmetric, analytic solution to the SU$(2)$ EYM system with a
positive cosmological constant (EYM$_{\Lambda}$) \cite{DH}.
Although, it turned out that the DH
solution is not really new, but gauge equivalent \cite{com}
to the cosmological solution of \cite{YH}, it triggered the work
presented in this paper. It is natural to expect that the cosmological
solution given in \cite{YH} is just the ground state of a discrete
family of static spherically symmetric solutions living on a
topological $3$-sphere. For reasons to be explained later, one
will, however, have to adjust the cosmological constant for each
member of the family. In my paper I will present the f\/irst excited
solution which I found numerically. As in the BK case, further
solutions were expected. In contrast to the analytic solution of
Hosotani or Ding and Hosoya, $g_{{}_{00}}$ is no longer constant
in my solution. Furthermore, I will prove some general properties
of the regular solutions, e.g. the interesting fact that solutions on $S^3$
can only exist for a non-vanishing cosmological constant.

The paper will be organized as follows: In chapter 2, I will
f\/irst show a derivation of the SU$(2)$ EYM$_{\Lambda}$ system
and the simplest way for obtaining the DH solution. This will
be followed by the above-mentioned general pro\-per\-ties of regular
solutions. In chapter 3, I will tell how I found the solutions
numerically, while the conclusions and the discussion can be
found in chapter~4. In conclusion, I will present a cosmological
embedding of the EYM$_{\Lambda}$ system which remarkably simplif\/ies
the consideration of stability of the DH solution. The idea can
be attributed to N. Straumann.
\newpage
%
\section{The EYM$_{\Lambda}$ dif\/ferential equations}
\subsection{Derivation of the dif\/ferential equation system}
We write the metric for a static, spherically symmetric
spacetime in the usual way:
\be \lb{met1}
g=-e^{2a}dt^{2}+e^{2b}dr^{2}
  +r^{2}(d\vartheta^{2}+\sin^{2}\!\vartheta\,d\varphi^{2})\:.
\ee
$a$ and $b$ are functions depending exclusively upon $r$.

We are only interested in YM fields that are purely magnetic.
According to Ref.~\cite{OS2}, we can determine the SU$(2)$ gauge
potential $A$ as
\be \lb{eich}
A=(-\tilde w \,\tau _1+w\,\tau _2)\,d\vartheta +(\tau _3\cot\vartheta -
w\,\tau _1 -\tilde w\,\tau _2 )\sin\!\vartheta \,d\varphi \:.
\ee
$w$, $\tilde{w}$ are dependent upon $r$, and $\tau_1$, $\tau_2$, $\tau_3$
are the spheric SU$(2)$ generators with $\vec{\tau}=\vec{\sigma}/2i$
($\vec{\sigma}$ being Pauli matrices). With a gauge
transformation we can furthermore make $\tilde w$ disappear,
with $A$ f\/inally assuming the following form:
\be \lb{eich3}
A={w\,\tau_2\,d\vartheta}+{(\tau_3\cos\vartheta-w\,\tau_1\sin\vartheta)\,d\varphi}\:.
\ee
This means we are looking for $a=a(r)$, $b=b(r)$ and $w=w(r)$.

With (\ref{eich3}), we can calculate the components of the
f\/ield strength tensor \mbox{$F=dA+A\wedge A$}
\be \lb{F}
F_{12}={w'\tau_2\,e^{-b}\,\frac{1}{r}}\:,\qquad
F_{13}={-w'\tau_1\,e^{-b}\,\frac{1}{r}}\:,\qquad
F_{23}=(w^2\!-1)\,\tau_3\,\frac{1}{r^2}\:.
\ee
We made use of the often employed orthonormal basis of 1-forms
\be \lb{for}
\theta^{0}={e^{a}dt}\:,\qquad
\theta^{1}={e^{b}dr}\:,\qquad
\theta^{2}={r\,d\vartheta}\:,\qquad
\theta^{3}={r\sin\!\vartheta\,d\varphi}\:,
\ee
and the notation $'\!\equiv {\partial _r}$. The def\/inition of the
YM-Lagrangian leads us to
\bd
L_{{}_{Y\!M}}={-\frac{1}{16\pi}\,tr(F_{\mu\nu}F^{\mu\nu})}\:,
\ed
\be \lb{lag}
L_{{}_{Y\!M}}={{\frac{1}{8\pi}\,{w'}^{2}\,\frac{e^{-2b}}{r^{2}}}\,+\,
      {\frac{1}{16\pi}\frac{(w^{2}-1)^{2}}{r^{4}}}}\:.
\ee

In the General Theory of Relativity (GR) we obtain the Einstein
f\/ield equations by varying the action
\be \lb{wir}
S=\int_{D}\Bigl[ -\frac{1}{16\pi G}\,R\,-\,\frac{1}{16\pi}\,
     tr(F_{\mu\nu}F^{\mu\nu}) \Bigr] \sqrt{-g}\,dx^4
\ee
with respect to the metric:
\be \lb{ein}
G^{\mu\nu}=8\pi G\,T^{\mu\nu}
\ee
\be \lb{eit}
T^{\mu\nu}=-\frac{1}{4\pi}\Bigl[ F_{\alpha\mu}F^{\alpha}_{\,\,\,\,\nu}\,-\,
    \frac{1}{2}\,g_{\mu\nu}\,(F|F) \Bigr]\:.
\ee
The components of the energy-momentum tensor become
\be \lb{eitk}
\begin{array}{ccl}
T_{00}&=& \frac{1}{8\pi}\,{w'}^{2}\,\frac{e^{-2b}}{r^{2}}\,+\,
          \frac{1}{16\pi}\frac{(w^{2}-1)^{2}}{r^{4}}\;,\\
T_{11}&=& \frac{1}{8\pi}\,{w'}^{2}\,\frac{e^{-2b}}{r^{2}}\,-\,
          \frac{1}{16\pi}\frac{(w^{2}-1)^{2}}{r^{4}}\;,\\
T_{22}=T_{33}&=& \frac{1}{16\pi}\frac{(w^{2}-1)^{2}}{r^{4}}
\end{array}
\ee
and the $G_{\mu\nu}$ relative to the basis (\ref{for}) are
\be \lb{eink}
\begin{array}{ccl}
G^0_{\,\,\,\, 0}&=& -\frac{1}{r^2}\,+\,
                    e^{-2b}(\frac{1}{r^2}-\frac{2b'}{r})\;,\\
G^1_{\,\,\,\, 1}&=& -\frac{1}{r^2}\,+\,
                    e^{-2b}(\frac{1}{r^2}+\frac{2a'}{r})\;,\\
G^2_{\,\,\,\, 2}=G^3_{\,\,\,\, 3}&=&
                  e^{-2b}({a'}^{2}-a'b'+a''+\frac{a'-b'}{r})\;.
\end{array}
\ee
All other $G_{\mu\nu}$ equal zero.

The spherically symmetric Einstein f\/ield equations with
$\Lambda$-term are
\be \lb{einl}
G_{\mu\nu}=8\pi G\,T_{\mu\nu}\,-\, g_{\mu\nu}\Lambda\:.
\ee
We insert (\ref{eitk}) and (\ref{eink}) into (\ref{einl}):
\be \lb{einl1}
-\frac{1}{r^2}\,+\,e^{-2b}\Bigl( \frac{1}{r^2}-\frac{2b'}{r} \Bigr)
=-G\,{w'}^{2}\,\frac{e^{-2b}}{r^2}\,
-\,\frac{G}{2}\frac{(w^2-1)^2}{r^4}\,-\,\Lambda\:,
\ee
\be \lb{einl2}
-\frac{1}{r^2}\,+\,e^{-2b}\Bigl( \frac{1}{r^2}+\frac{2a'}{r} \Bigr)
=G\,{w'}^{2}\,\frac{e^{-2b}}{r^2}\,
-\,\frac{G}{2}\frac{(w^2-1)^2}{r^4}\,-\,\Lambda\:.
\ee

For the YM equations, we can operate in a similar way as in
obtaining in the GR the f\/ield equations from a variational
principle. Here, the action is
\be \lb{wym}
S_{{}_{Y\!M}}=\int\!\!{\cal{L}}_{{}_{Y\!M}}\eta\quad.
\ee
If we now vary with respect to the gauge potential $A$,
we obtain
\be \lb{ymg1}
D\ast F=\,0\quad.
\ee
In addition, we can also prove the identity
\be \lb{ymg2}
DF=D(dA+A\wedge\!A)=\,0\:.
\ee
The best way to do this, is to use $D\,.=d\,.+[A,\,.\,]$\,, as well as
\mbox{$[B,C]=B\wedge C-(-1)^{{}^{pq}}C\wedge B$}\quad for\quad
$B\in \Lambda ^{p}(M)$, $C\in \Lambda ^{q}(M)$.
The equations (\ref{ymg1}) and (\ref{ymg2}) are the known
YM equations.

But now, back to our problem. We write the equation (\ref{ymg1})
for our $A$ (\ref{eich3}) and our basis $\{\theta ^\mu\}$
(\ref{for}) in full:
\be \lb{bw}
e^{-2b}\,\frac{w''+w'(a'-b')}{r}\,-\,\frac{(w^2-1)w}{r^3}=\,0\:.
\ee
(\ref{bw}) is the equation of motion of $w$. The equations (\ref{einl1}),
(\ref{einl2}) and (\ref{bw}) constitute a complete set of equations
for the static spherically symmetric SU$(2)$ EYM$_{\Lambda}$ system.
Now, we have three dif\/ferential equations for $a(r)$, $b(r)$ and $w(r)$. We
write them down for $m(r)$ and $\delta (r)$ instead of for
$a(r)$ and $b(r)$, with
\begin{eqnarray}
\delta (r) &:=& -a(r)-b(r)\:, \lb{del} \\
e^{-2b(r)} &:=& 1- \frac{2\,m(r)}{r}\:, \lb{eb} \\
N(r) &:=& e^{-2b(r)} = 1- \frac{2\,m(r)}{r}\:, \lb{N} \\
S(r) &:=& e^{-\delta (r)}\:.\lb{S}
\end{eqnarray}
{}From (\ref{einl1})$-$(\ref{einl2}) and $\delta '=-(a'+b')$,
we obtain
\be \lb{delp}
\delta '=-\,G\,\frac{{w'}^{2}}{r}\quad.
\ee
Then we dif\/ferentiate (\ref{eb}) with respect to $r$ and insert
the result into (\ref{einl1}) (with regard to (\ref{N})\,):
\be \lb{mp}
m'=\frac{G}{2}\,{w'}^{2}\,N(r)\,+\,
   \frac{G}{4}\,\frac{(w^2-1)^2}{r^2}\,+\,\frac{\Lambda}{2}\,r^2\:.
\ee
We set $V\!:=\frac{(w^2-1)^2}{2r^2}$, note that $N'=\frac{2m}{r^2}-
\frac{2m'}{r}$, multiply both sides by 2 and obtain
\bd
2m'=G\,(V\!+\!N{w'}^{2})+\Lambda\,r^2\:.
\ed
With $2m'=\frac{2m}{r}-rN'$ and $N-1=-\frac{2m}{r}$ the whole turns
into
\be \lb{Np}
-r\,N'=G\,(V\!+\!N{w'}^{2})+(N-1)+\Lambda\,r^2\:.
\ee
We write (\ref{bw}) in a slightly dif\/ferent way
\be \lb{bw1}
e^{-2b}\,w''\,+\,e^{-2b}\,w'\,(a'-b')\,-\,\frac{w^2-1}{r^2}\,w=0\:.
\ee
Then, we consider the expression $\Bigl( Ne^{-\delta}w'\Bigr) '$.
Writing it out and multiplying by $e^{\delta}$ gives
\be \lb{bw2}
\frac{1}{e^{-\delta}}\,\Bigl( Ne^{-\delta}w'\Bigr) '=\,
e^{-2b}\,w'\,(a'\!-b')\,+\,e^{-2b}\,w''\;.
\ee
Comparison of (\ref{bw2}) with (\ref{bw1}) gives us:
\be \lb{bw3}
\Bigl( Ne^{-\delta}w'\Bigr) '=\frac{w^2-1}{r^2}\,w\,e^{-\delta}\;.
\ee
The def\/inition of $V$ leads to:
\bd
\frac{1}{2}\,\frac{\partial V}{\partial w}=\frac{w(w^2-1)}{r^2}\:,\qquad
\frac{\partial V}{\partial w}\equiv V_{{}_{'}{w}}\:.
\ed
Thus, (\ref{bw3}) becomes:
\be \lb{bw4}
\Bigl( Ne^{-\delta}w'\Bigr) '=\frac{1}{2}\,V_{{}_{'}{w}}\,e^{-\delta}\:.
\ee
We summarize the results:
\begin{eqnarray}
-r\,N' &=& G\,(V\!+\!N{w'}^2)\,+\,(N-1)\,+\,\Lambda\,r^2 \lb{e1} \\
-r\,\delta ' &=& G\,{w'}^2 \lb{e2} \\
\Bigl( Ne^{-\delta}w'\Bigr) ' &=& \frac{1}{2}\,V_{{}_{'}{w}}\,e^{-\delta}\:,
\lb{ym}
\end{eqnarray}
with
\begin{eqnarray*}
V &=& \frac{(1-w^2)^2}{2r^2} \\
\frac{1}{2}\,V_{{}_{'}{w}} &=& \frac{w(w^2-1)}{r^2}\;.
\end{eqnarray*}
(\ref{e1}) - (\ref{ym}) constitute for the functions $N(r)$,
$\delta (r)$, $w(r)$ our complete system of dif\/ferential equations
of the SU$(2)$ EYM$_{\Lambda}$ system.


\subsection{An analytical solution to the system}
In 1993, S. Ding and A. Hosoya presented an analytical
solution to the system (\ref{e1}) - (\ref{ym}) \cite{DH}.
I will give within the framework of our formulas another
derivation.

We make the ansatz $S(r)\equiv N(r)^{\alpha}$, $\alpha =-\frac12$.
This, along with (\ref{S}), gives us
\be \lb{delp1}
\delta '=\frac{1}{2}\,\frac{N'}{N}\;.
\ee
Our system (\ref{e1}) - (\ref{ym}) thus becomes
\begin{eqnarray}
-r\,N' &=& N\!-1\,+\,G(V\!+\!N{w'}^{2})+\Lambda\,r^2 \lb{e3} \:, \\
-r\,N' &=& 2\,G\,N\,{w'}^{2} \lb{e4} \:, \\
\Bigl( \sqrt{N}\,w' \Bigr) ' &=& \frac{1}{2\sqrt{N}}\,V_{{}_{'}{w}}
\lb{ym1} \;.
\end{eqnarray}
{}From (\ref{e3})$-$(\ref{e4}), we obtain
\bd
GN{w'}^2=N\!-\!1\,+\,G\,V\,+\,\Lambda\,r^2
\ed
or
\be \lb{e5}
G\Bigl( \sqrt{N}w' \Bigr) ^2=N\!-\!1\,+\,G\,V\,+\,\Lambda\,r^2\;.
\ee
We calculate the derivative of (\ref{e5}) with respect to $r$:
\be \lb{e6}
2\,G\,\Bigl( \sqrt{N}w'\Bigr) \Bigl(\sqrt{N}w'\Bigr) '=
   \Bigl( N\!-\!1\,+\,G\,V\,+\,\Lambda\,r^2\Bigr) '\:.
\ee
On the other hand, (\ref{ym1}) leads to
\be \lb{e7}
2\,G\,\Bigl( \sqrt{N}w'\Bigr) \Bigl(\sqrt{N}w'\Bigr) '=
   G\,V_{{}_{'}{w}}\,w'\:.
\ee
We compare (\ref{e6}) with (\ref{e7}), bearing in mind that
$V'\!=V_{{}_{'}{w}}\,w'+V_{{}_{'}{r}}=$
\mbox{$V_{{}_{'}{w}}\,w'-\frac{2}{r}\,V$}.
Thus, we obtain
\be \lb{sol1}
N'=2\,G\frac{V}{r}\,-\,2\,\Lambda\,r\:.
\ee
In (\ref{sol1}), we eliminate $N'$ with (\ref{e4})
\be \lb{sol2}
N{w'}^{2}\,+\,V=\frac{\Lambda}{G}\,r^2\:.
\ee
We insert this into (\ref{e3}) and obtain
\be \lb{sol3}
N'=-\frac{N}{r}\,+\,\Bigl( \frac{1}{r}-2\Lambda r\Bigr) \:.
\ee
We clearly see that (\ref{sol3}) is linear.

The corresponding solution is:
\bd
N=\frac{c}{r}\,+\,\Bigl( 1-\frac{2}{3}\Lambda r^2\Bigr)\:.
\ed
$c$ is an integration constant. With the condition $N(0)<\infty$,
$c$ equals zero. Thus:
\be \lb{sol4}
N=1-\frac{2}{3}\Lambda r^2\:.
\ee
We dif\/ferentiate (\ref{sol4}): $N'=-\frac43\,\Lambda\,r$,
insert this into (\ref{sol1}) and solve for $w$:
\be \lb{sol5}
w^2=1-\sqrt{\frac{2}{3}\frac{\Lambda}{G}}\,r^2\:.
\ee
If we insert (\ref{sol4}) and (\ref{sol5}) into (\ref{e4}),
we see that
\be \lb{sol6}
1=\frac{N}{w^2}\:.
\ee
According to (\ref{sol6}), we equate (\ref{sol4}) with (\ref{sol5})
and obtain a quadratic equation in $\Lambda$:
\bd
\Lambda\left( \frac{2G}{3}\Lambda -1\right) =0\:.
\ed
The sensible solution is $\Lambda =\frac{3}{2G}$ which results
f\/inally in the DH solution
\be \lb{sol}
w^2=N=1-\frac{r^2}{G}\:.
\ee
%
\subsection{Interpretation of the solution}
Our ansatz has been: $S=N^{-1/2}$, which means that $e^{2a}\!=1$.
With this and with (\ref{sol}), the metric (\ref{met1}) becomes
\be \lb{met2}
g=-dt^2\,+\,\frac{1}{1-\frac{r^2}{G}}\,dr^2\,+\,r^2d\Omega ^2\:.
\ee
We want to replace $r$ by
\bd
r=\frac{u}{1+\frac{u^2}{4G}}
\ed
and thus re-write our metric:
\be \lb{met3}
g=-dt^2\,+\,\frac{1}{\left( 1+\frac{u^2}{4G}\right) ^2}\,du^2\,+\,
     \frac{u^2}{\left( 1+\frac{u^2}{4G}\right) ^2}\,d\Omega ^2\:.
\ee
Thus, our solution (\ref{sol}) describes an Einstein universe \cite{NS}.

F\/inally, we observe
\be \lb{rho}
d\varrho =\frac{dr}{\sqrt{1-\frac{r^2}{G}}}\:.
\ee
This can be solved:
\bd
r=\sqrt{G}\,\sin\!\left( \frac{\varrho}{\sqrt{G}}+c\right)\:.
\ed
c is an integration constant. If we require that $r(\varrho =0)=0$,
c becomes zero. Thus:
\be \lb{r}
r=\sqrt{G}\,\sin\!\frac{\varrho}{\sqrt{G}}\:.
\ee
Now, the metric (\ref{met2}) can be written as follows:
\be \lb{met4}
g=-dt^2\,+\,d\varrho ^2\,+\,G\,\sin ^2\!\left( \frac{\varrho}{\sqrt{G}}
    \right)\,d\Omega ^2\:.
\ee
This means that the spacetime manifold is ${\Bbb R}\times
{\Bbb S}^3$, $\varrho$ being an angular coordinate of ${\Bbb S}^3$.
%
\subsection{The EYM$_{\Lambda}$ equations for $L$, $R$, $w$}
We set up the EYM$_{\Lambda}$ equations again, this time for the functions
$L$, $R$, $w$ of the coordinate $\varrho$. Our metric
now reads like this:
\be \lb{met}
g=-L^2dt^2\!+d\varrho ^2\!+R^2\!(\varrho )\,d\Omega ^2\:,
\ee
\bd
R\equiv r\;, \quad L=\sqrt{N} S\: .
\ed
We use our old EYM$_{\Lambda}$ equations (\ref{e1}) - (\ref{ym}) and the
relation
\be \lb{bed}
\frac{dr}{d\varrho }\equiv \sqrt{N}=\dot R\:\:.
\ee
We use the notation:
\bd
'\equiv \frac{d}{dr}\:,\qquad \dot {}\, \equiv \frac{d}{d\varrho }\:.
\ed
First, we need
\be \lb{zu1}
N'=\left( {\dot R}^{\,2} \right) '=\left( {\dot R}^{\,2}\right) ^{
\cdot}\,\frac{d\varrho }{dr}=2\ddot R\:,
\ee
\be \lb{zu2}
w'=\dot w\,\frac{d\varrho }{dr}=\frac{\dot w}{\dot R}\quad .
\ee
We insert that into (\ref{e1})
\be \lb{R1}
\ddot R=-\frac{1}{2R}\,\left[ G\,(V\!+{\dot w}^{2})+
({\dot R}^{\,2}\!-1)+\Lambda R^2\right] \:.
\ee

Next, we need
\be \lb{S1}
S=\frac{L}{\dot R}=e^{-\delta}\:.
\ee
Its derivative results in
\be \lb{delp2}
-\delta '=\frac{\dot L}{L\dot R}\,-\,\frac{\ddot R}{{\dot R}^{\,2}}\:.
\ee
(\ref{delp2}) and (\ref{zu2}) turn (\ref{e2}) into
\bd
R\,\frac{\dot L}{L\dot R}\,-\,R\,\frac{\ddot R}{{\dot R}^{\,2}}=
G\,\frac{{\dot w}^{2}}{{\dot R}^{\,2}}
\ed
and solved for $\dot L$
\be \lb{L1}
\dot L=\frac{L\ddot R}{\dot R}\,+\,G\,\frac{{\dot w}^{2} L}{R\dot R}\quad .
\ee

With the help of (\ref{bed}), (\ref{zu2}) and (\ref{S1}), (\ref{ym})
becomes
\be \lb{w1}
\dot L\,\dot w\,+\,L\,\ddot w=\frac{1}{2}\,V_{{}_{'}{w}}\,L\:.
\ee
By inserting (\ref{L1}) into (\ref{w1}), we can eliminate $L$:
\be \lb{w2}
\ddot w =\frac{1}{2}\,V_{{}_{'}{w}}\,-\,
\frac{\dot w}{\dot R}\left( G\,\frac{{\dot w}^{2}}{R}+\ddot R\right) \:.
\ee
And with (\ref{R1}), $\ddot R$ also disappears:
\be \lb{w3}
\ddot w =\frac{1}{2}\,V_{{}_{'}{w}}\,+\,\frac{\dot w}{2R\dot R}
\left[ G\,(V\!-{\dot w}^{2})+({\dot R}^{2}\!-1)+\Lambda R^2\right] \:.
\ee
Summarizing:
\begin{eqnarray}
\ddot R & = & -\frac{1}{2R}\,\left[ G\,(V\!+{\dot w}^{2})+
({\dot R}^{\,2}\!-1)+\Lambda R^2\right] \lb{R2} \\
\ddot w & = & \frac{1}{2}\,V_{{}_{'}{w}}\,+\,\frac{\dot w}{2R\dot R}
\left[ G\,(V\!-{\dot w}^{2})+({\dot R}^{2}\!-1)+\Lambda R^2\right] \lb{w4} \\
\dot L & = & \frac{L\ddot R}{\dot R}\,+\,G\,\frac{{\dot w}^{2} L}{R\dot R}
\lb{L2} \quad .
\end{eqnarray}

We know from our special solution (\ref{r}) that $R(\varrho )$ becomes
maximal at the equator, i.e.:
\bd
\left. \dot R(\varrho) \right| _{\varrho =\sqrt{G}\,\frac{\pi}2}=\,0\:.
\ed
To regularize our equations (\ref{R2}) - (\ref{L2}), we introduce
a new function that upon considering the equations is perfectly
obvious:
\be \lb{gam1}
\gamma (\varrho ):=\frac{1}{R\dot R}\,\left[ G\,(V\!-{\dot w}^{2})+({\dot
R}^{\,2}\!-1)+\Lambda R^2\right]\:.
\ee
Thus, (\ref{R2}), (\ref{w4}) and (\ref{L2}) become:
\begin{eqnarray*}
\ddot R & = & -\,\frac12\,\dot R\,\gamma\,-\,G\,\frac{{\dot w}^2}{R}\:, \\
\ddot w & = & \frac12\,V_{{}_{'}{w}}\,+\,\frac12\,\dot w\,\gamma\:, \\
\dot L & = & -\,\frac12\,L\,\gamma\:.
\end{eqnarray*}
We also need $\dot{\gamma}$. After a longer, but not at all
dif\/f\/icult calculation, we f\/ind that
\bd
\dot \gamma =\frac12\,\gamma ^2\,-\,\frac{2}{R^2}\,\left[ 2G\,V+
{\dot R}^{\,2}-1\right]\:.
\ed
Our f\/inal system now looks like this:
\begin{eqnarray}
\ddot w & = & \frac12\,V_{{}_{'}{w}}\,+\,\frac12\,\dot w\,\gamma \lb{w}\:, \\
\ddot R & = & -\,\frac12\,\dot R\,\gamma\,-\,G\,\frac{{\dot w}^2}{R}
\lb{R}\:,\\
\dot \gamma & = & \frac12\,\gamma ^2\,-\,\frac{2}{R^2}\,\left[ 2G\,V+
{\dot R}^{\,2}-1\right] \lb{gamp}\:, \\
\dot L & = & -\,\frac12\,L\,\gamma\lb{Lp}\quad ,
\end{eqnarray}
with
\begin{eqnarray}
\gamma & = & \frac{1}{R\dot R}\,\left[ G\,(V\!-{\dot w}^{2})+({\dot
R}^{\,2}\!-1)+\Lambda R^{\,2}\right] \lb{gam}\:, \\
V & = & \frac{(w^2-1)^2}{2R^2} \lb{pot}\:, \\
\frac12\,V_{{}_{'}{w}} & = & \frac{w(w^2-1)}{R^2}\lb{dpot}\quad .
\end{eqnarray}
{}From now on, we will work only with this system.

In conclusion, I will show how the solution of Ding \& Hosoya
\cite{DH} looks in these coordinates. We can f\/ind them
quickly with the help of (\ref{r}) and (\ref{sol}):
\begin{eqnarray}
R & = & \sqrt{G}\,\sin\!\left(\frac{\varrho}{\sqrt{G}}\right)\lb{RDH}\\
w & = & \cos\!\left( \frac{\varrho}{\sqrt{G}}\right)\lb{wDH}\\
L & = & \dot R S=\sqrt{\frac{N}{N}}=1\lb{LDH}\\
\gamma & = & 0\lb{gDH}\quad ,
\end{eqnarray}
\bd
\varrho \in \left[ 0,\sqrt{G}\pi\right]\:.
\ed
%
\subsection{Expansions}\lb{exp}
We already stated that in the case $\varrho =0$ we require that
$R=0$. Furthermore, in $\gamma$ (\ref{gam}) we encounter the term
$\Lambda R^{\,2}$. As the spacetime manifold has the topology of $S^3$,
R equals zero again at a certain $\varrho$. We will call
this $\varrho _{{}_0}$. Thus, we have
\bd
R(0)=0\:,\qquad
R({\varrho}_{{}_0})=0\:.
\ed
For the solution (\ref{RDH}) - (\ref{gDH}), $\varrho _{{}_0}=
\sqrt{G}\pi$. Unfortunately, in our system \mbox{(\ref{w}) - (\ref{dpot})}
$R$ also appeares in the denominator; therefore, we have
to expand our functions $R$, $w$, $\gamma$, $L$ at
$\varrho =0$ and $\varrho =\varrho _{{}_0}$.

Let us f\/irst take a look at (\ref{N}). As we want regular solutions,
$m(r)\to 0$ must hold for $r\to 0$, and thus $N(r)\to 1$.
According to (\ref{bed}), that means that
\bd
{\dot R}^{\,2} \!\to 1\qquad
(\varrho \to 0\,\,\mbox{or}\,\,\varrho \to {\varrho}_{{}_0})\:.
\ed
If $R\to 0$, then $w\to\pm 1$, since $V$ otherwise becomes singular,
which is obvious from (\ref{pot}). Equation (\ref{R}) gives us
\bd
R\,\ddot R=-\,\frac12 \,R\,\dot R \,\gamma \,-\,G\,{\dot w}^2\:.
\ed
We see immediately that $\dot w\to 0$, if $\varrho\to 0$ or
$\varrho\to\varrho _{{}_0}$. This corresponds to the observation
made by BK \cite{BK}, which says that the solutions become
asymptotically to $w=\pm 1$, since $\dot w$ must be zero there.

With enough patience in dif\/ferentiation and applying the rule of
Bernoulli-de l'H\^opital several times, the equations
(\ref{w}) - (\ref{dpot}) provide us the important conditions:
\bd
\left.
\begin{array}{c}
\ddot R \to 0 \\
R^{{}^{(3)}}\!\to -\frac{G{\ddot w}^2}{2\dot R}-\frac{\Lambda}{3\dot R}
\end{array} \right \}
\varrho \to 0\;,\,\varrho \to {\varrho}_{{}_0}\:.
\ed
$R^{{}^{(3)}}$ and $\ddot w$ can be any values. We set
$R^{{}^{(3)}}:=\mp 2c, c>0$ and $\ddot w:=-2b, b>0$. We summarize
our results again:
\bd
\left.
\begin{array}{ccc}
R &\to& 0 \\ \dot R &\to& 1 \\ \ddot R &\to& 0 \\ R^{{}^{(3)}} &\to& -2c \\
w &\to& 1 \\ \dot w &\to& 0 \\ \ddot w &\to& -2b
\end{array} \right \}
\varrho \to 0 \qquad \left.
\begin{array}{ccc}
R &\to& 0 \\ \dot R &\to& -1 \\ \ddot R &\to& 0 \\ R^{{}^{(3)}} &\to& +2c \\
w &\to& \pm 1 \\ \dot w &\to& 0 \\ \ddot w &\to& -2b
\end{array} \right \}
\varrho \to {\varrho}_{{}_0}
\ed
These conditions lead to:
\begin{eqnarray}
c&=&G\,b^2\,+\,\frac{\Lambda}{6} \lb{c} \:, \\
R & = & \varrho \,-\,\frac{c}{3}\,{\varrho}^3 \lb{Re1}\:, \\
w & = & 1\,-\,b\,\varrho ^2 \lb{we1}\:, \\
R & = &({\varrho}_0-\varrho )\,-\,\frac{c}{3}\,({\varrho}_{{}_0}-\varrho )^3
\lb{Re2}\:, \\
w & = & \mp 1\,\pm\,b\,({\varrho}_{{}_0}-\varrho )^2\lb{we2}\:;
\end{eqnarray}
$\pm$ in the formula (\ref{we2}) depends upon whether $w$ has an even
or an odd number of zeros. (\ref{c}) - (\ref{we2}) lead to the
expansion of $\gamma$:
\begin{eqnarray}
\gamma & = & \left( -4Gb^2+\frac23 \Lambda \right)\varrho\:,\quad\varrho\to 0
\lb{ge1}\\
\gamma & = & \left( 4Gb^2-\frac23 \Lambda\right) (\varrho -{\varrho}_{{}_0})\:,
\quad\varrho\to {\varrho}_{{}_0}\lb{ge2}\:.
\end{eqnarray}
Now we can insert (\ref{ge1}) and (\ref{ge2}) into (\ref{Lp}), in order
to obtain $L$:
\begin{eqnarray}
L & = & 1+\left( Gb^2-\frac{\Lambda}{6}\right) \varrho ^2\:,
\quad \varrho \to 0 \lb{Le1}\\
L & = & 1+\left( Gb^2-\frac{\Lambda}{6}\right) (\varrho _{{}_0} -\varrho )^2\:,
\quad \varrho \to \varrho _{{}_0}\lb{Le2}\:.
\end{eqnarray}

As an example, we can expand (\ref{RDH}) and (\ref{wDH}) and compare
them with (\ref{c}) - (\ref{we1}). Thus, we obtain $b=0.25$ and
$\Lambda =0.75$, if we take $G=2$. When solving numerically,
we need those values.
%
\subsection{Properties of the dif\/ferential equation system}
We have seen in Sec.~2.5. that the function $w$ starts at $1$ and
ends at $\pm 1$. We will now show that $w$ cannot become higher
than $+1$ or lower than $-1$ within the interval $]0,\varrho _{{}_0}[$,
as it would otherwise never return and diverge.

Furthermore, we will see that any solution with $\Lambda\equiv 0$
is trivial, i.e. $R\equiv 0$, $w\equiv\pm 1$.
Then I will provide a proof that $R$ can have only maxima and no
minima within the interval $[0,\varrho _{{}_0}]$.
In conclusion, we will recognize that at the points where $R$ has a
saddle point\footnote{point where the f\/irst and second derivative
vanishes \cite{St}}, $w$ will have an extremum. All these
facts combined allow for a qualitative idea of the solutions.

\begin{proposition}\lb{prop1}
For any solution $w$, the following applies:
\bd
|w(\varrho )|\leq 1\:,\quad \forall \varrho\in\, ]0,\varrho _{{}_0}[\:.
\ed
\end{proposition}
{\em{Proof:} } Assume the contrary, i.e., that $w$
is larger than $+1$ for certain \mbox{$\varrho\!\in\, ]0,\varrho _{{}_0}[$.}
In order that the behavior at $\varrho =0$ and $\varrho =
\varrho _{{}_0}$ holds, $w$ must have a maximum in that
range, where it is larger than $+1$. That means that there is a
$\tilde{\varrho}\!\in\, ]0,\varrho _{{}_0}[$ where the following
applies:
\bd
w(\tilde{\varrho})>1\:,\quad \dot w(\tilde{\varrho})=0\:,
\quad \ddot w(\tilde{\varrho})\le0\:.
\ed
Now we use equation (\ref{w}):
\bd
\ddot w=\frac{w\,(w^2\!-1)}{R^2}\,+\,\frac12\,\dot w\,\gamma\:.
\ed
R does not vanish at $\tilde{\varrho}$ and thus, we have
\bd
\ddot w(\tilde{\varrho})=\left. \frac{w\,(w^2\!-1)}{R^2}\right|
_{\tilde{\varrho}}\:,
\ed
implying
\bd
\ddot w(\tilde{\varrho})>0\:.
\ed
This is a contradiction to the assumption. We can proof the case
$-1$ ana\-logically.
\begin{flushright}
{\em q.e.d.}
\end{flushright}
In order to prove that nontrivial solutions must
have $\Lambda\ne 0$, we need several tools.
\begin{proposition}\lb{prop2}
Every nontrivial solution has at least one zero of $w$.
Thus, there has to be a $\varrho ^{*}\!\in\,]0,\varrho _{{}_0}[$
with $w(\varrho ^{*})=0$.
\end{proposition}
{\em{Proof:} } We make again a counterassumption and say we have a
$w$ for which the following holds:
\bd
w(\varrho ^{*})\ne 0\:,\quad \forall \varrho ^{*}\!\in\,]0,\varrho _{{}_0}[\:.
\ed
Without loss of generality we can chose $w(0)=+1$. Then $w$
must have a minimum in order to fulf\/ill $w(\varrho _{{}_0})=+1$.
It cannot be equal to $-1$, because in that case $w$ would have a zero.
Therefore, there is a $\hat{\varrho} \!\in\,]0,\varrho _{{}_0}[$
for which
\begin{eqnarray*}
w(\hat{\varrho} )>0\:,&\quad & w(\hat{\varrho} )<1\quad (\mbox{follows from
Prop.~\ref{prop1}})\\
\dot w(\hat{\varrho} )=0\:,&\quad & \ddot w(\hat{\varrho} )>0\:.
\end{eqnarray*}
Taking a look back at equation (\ref{w}),
\bd
\ddot w(\hat{\varrho} )=\frac{w(\hat{\varrho} )\left[ w^2(\hat{\varrho}
)-1\right]}
{{R}^{\,2}(\hat{\varrho} )}\:,
\ed
we see that \quad $\ddot w(\hat{\varrho} )<0$.\\
This is a contradiction to the assumption. We can analogiously
treat the case $w(0)=-1$.
\begin{flushright}
{\em q.e.d.}
\end{flushright}

For the ground state with \mbox{$w(0)=+1$,} \mbox{$w(\varrho _{{}_0})=-1$},
there is exactly one zero.

\begin{proposition}\lb{prop3}
If there is a $\tilde{\varrho}\!\in\,]0,\varrho _{{}_0}[$ with
$\gamma (\tilde{\varrho})=0$, then $\dot{\gamma}(\tilde{\varrho})<0$,
if we presuppose $\Lambda = 0$.
\end{proposition}
{\em{Proof:} } We chose a $\tilde{\varrho}$ with $\gamma (\tilde{\varrho})=0$.
Equation (\ref{gam}) gives us
\bd
G\,V(\tilde{\varrho})-G\,{\dot w}^{2}(\tilde{\varrho})
+{\dot R}^{\,2}(\tilde{\varrho})-1=0\:.
\ed
We solve this for $G\,V$ and insert the result into (\ref{gamp}).
Thus, we obtain
\bd
\dot{\gamma}(\tilde{\varrho})=
-\frac{2\,G}{R^{\,2}(\tilde{\varrho})}\,\left[ \,2G\,{\dot w}^2
(\tilde{\varrho})-{\dot R}^{\,2}(\tilde{\varrho})+1\,\right] \:.
\ed
Since ${\dot w}^2$ and $R^{\,2}$ are always larger than zero
and $0\le {\dot R}^{\,2}\le 1$\,,\,$\dot{\gamma}(\tilde{\varrho})<0$.
\nopagebreak[1]
\begin{flushright}
{\em q.e.d.}
\end{flushright}
\begin{proposition}\lb{prop4}
For every solution with $w(0)=+1$ and $w(\varrho _{{}_0})=-1$
and $\Lambda = 0$ $\gamma$ must vanish identically.
\end{proposition}
{\em{Proof:} } The equations (\ref{ge1}) and (\ref{ge2}) and the
condition $\Lambda = 0$ lead us to
\begin{eqnarray*}
\dot\gamma (\varrho =0)&<&0 \\
\dot\gamma (\varrho =\varrho _{{}_0})&<&0\:.
\end{eqnarray*}
However, since $\gamma (0)=\gamma (\varrho _{{}_0})=0$ still
applies, $\gamma$ must cross the abscissa at least once;
e.g. at $\tilde\varrho$, where $\dot\gamma (\tilde\varrho )>0$.
This, however, is a contradiction to Proposition~\ref{prop3},
which says that $\dot\gamma (\tilde\varrho )<0$. Therefore,
only $\gamma\equiv 0$ can apply.
\begin{flushright}
{\em q.e.d.}
\end{flushright}
\begin{lemma}\lb{lem1}
Every solution with $\Lambda = 0$ is trivial,
i.e. $R\equiv 0$, $w\equiv\pm 1$.
\end{lemma}
{\em{Proof:} } We know from Proposition~\ref{prop4} that
\bd
\gamma \equiv 0 \quad \forall \varrho \!\in
[0,\varrho _{{}_0}]\:.
\ed
Together with (\ref{gamp}) and (\ref{gam}), this leads to
\begin{eqnarray*}
2G\,V+{\dot R}^{\,2}-1&=&0\:, \\
G\,V-G\,{\dot w}^2+{\dot R}^{\,2}-1&=&0\:.
\end{eqnarray*}
We subtract the two equations from each other and obtain:
\bd
{\dot w}^2 =-V\:.
\ed
Since $V\ge 0$, it must be:
\bd
\dot w =0\;,\quad\forall\varrho\!\in [0,\varrho _{{}_0}]\:.
\ed
Therefore, $w$ must be proportional to a constant. In order to
keep $V$ regular for $R(0)=0$, only $w\equiv\pm 1$ can apply. \\
With (\ref{R}), we can see that $\ddot R\equiv 0$. Therefore, R
should equal zero or be proportional to $\varrho$. However, if it
were proportional to $\varrho$, it could never equal zero at
$\varrho _{{}_0}$. Therefore: $R\equiv 0$.
\begin{flushright}
{\em q.e.d.}
\end{flushright}
Of course, Lemma~\ref{lem1} only applies if we assume that
$R$ equals zero again for a f\/inite value of $\varrho _{{}_0}$.
But we had already assumed this in Sec.~\ref{exp}.

Now we will take a look at the properties of $R$.
\begin{proposition}\lb{prop5}
The maximum of $R$ is positive, the minimum negative.
\end{proposition}
{\em{Proof:} }
\begin{description}
\item[Maximum:]
$\exists\varrho ^{\ast}\!\in\,]0,\varrho _{{}_0}[$\,, so that
\bd
\dot R(\varrho ^{\ast})=0\;,\quad \ddot R(\varrho ^{\ast})<0\:.
\ed
We insert into equation (\ref{R}):
\bd
\ddot R(\varrho ^{\ast})=-G\,\frac{{\dot w}^2(\varrho ^{\ast})}{R(\varrho
^{\ast})}\:.
\ed
${\dot w}^2$ is always positive\,; therefore, $R(\varrho ^{\ast})>0$.
\item[Minimum:]
$\exists\varrho ^{\ast}\!\in\,]0,\varrho _{{}_0}[$\,, so that
\bd
\dot R(\varrho ^{\ast})=0\;,\quad \ddot R(\varrho ^{\ast})>0\:.
\ed
With the same argument as above, we show that: $R(\varrho ^{\ast})<0$.
\end{description}
\begin{flushright}
{\em q.e.d.}
\end{flushright}
Minima only exist when the radius $R$ at these points is negative.
In order to fulf\/ill the boundary conditions $\dot R(0)=+1$,
$\dot R(\varrho _{{}_0})=-1$, $R$ must in this case have two
maxima or two zeros. And for every additional minimum there must
be an additional maximum. In other words:
\begin{center}
{\bf {Number of maxima $=$ Number of minima $+1$.} }
\end{center}
If we exclude negative values for all regular solutions to
$R$, we obtain the information that all nontrivial, regular
solutions to $R$ have just one maximum.

We also know something about the saddle points.

\begin{proposition}\lb{prop6}
If $R$ has saddle points, then $w$ has an
extremum at this same point.
\end{proposition}
{\em{Proof:} } If $R$ has a saddle point or a f\/lat point, the
following applies:
\bd
\exists \tilde{\varrho}\!\in\, ]0,\varrho _{{}_0}[,\quad
\mbox{so that}\,\,\dot R(\tilde{\varrho})=0\;, \ddot R(\tilde{\varrho})=0\:.
\ed
We insert into (\ref{R}) and obtain
\bd
\ddot R(\tilde{\varrho})=-\frac12\,\dot
R(\tilde{\varrho})\,\gamma(\tilde{\varrho})
-G\,\frac{{\dot w}^2(\tilde{\varrho})}{R(\tilde{\varrho})}=
-\,G\,\frac{{\dot w}^2(\tilde{\varrho})}{R(\tilde{\varrho})}=0\:.
\ed
Since $R(\tilde{\varrho})\ne 0$, $\dot w(\tilde{\varrho})$
must be equal to zero.
\begin{flushright}
{\em q.e.d.}
\end{flushright}
To provide an overall view, we summarize our results:
\begin{itemize}
\item All regular solutions of $w$ lie between $+1$ and $-1$.
\item Nontrivial, regular solutions exist only for $\Lambda\ne 0$.
\item If we exclude negative values of $R$, it has only one maximum.
\item $R$ can have saddle points; at these points,
$w$ has an extremum.
\end{itemize}
All these results are of particular interest when we know
whether the solutions have any symmetry. For example, a
ref\/lection at the equator would be nice. \\
We will now consider the ref\/lection at an axis and the point
ref\/lection $(G=1)$. $\varrho$ then runs from $0$ to
$\varrho _{{}_0}$. Because of the symmetry the interval is
divided into two new ones. We don't want to assess where this
is, but chose arbitrarily:
\begin{eqnarray}
&\varrho\in [0,\varrho _{{}_0}]=:I& \nonumber\\
&\varrho ':=\frac{\varrho}{\varepsilon}\;,\quad
\varrho '':=\varrho _{{}_0}-\frac{\varepsilon -1}{\varepsilon}\varrho\;,
\quad \varepsilon\in{\Bbb R}_{+}\!\setminus\!\{0\}&\lb{rho1}\:.
\end{eqnarray}
This leads us to
\begin{eqnarray*}
&\varrho '\in\left[ 0,\frac{\varrho _{{}_0}}{\varepsilon}\right] =:I_1\;,
\quad \varrho ''\in\left[\varrho _{{}_0},\frac{\varrho _{{}_0}}
{\varepsilon}\right] =:I_2& \\
&I_1\cup I_2=I\:.&
\end{eqnarray*}
Our symmetries shall be:
\begin{eqnarray}
1)\quad \left.
\begin{array}{rcl}
R(\varrho ')&=&R(\varrho '')\\
w(\varrho ')&=&w(\varrho '')
\end{array} \right\}\quad
\mbox{(ref\/lection)}\lb{sym1}\\
2)\quad \left.
\begin{array}{rcl}
R(\varrho ')&=&R(\varrho '')\quad\mbox{(ref\/lection)}\\
w(\varrho ')&=&-w(\varrho '')\quad\mbox{(point ref\/lection)}
\end{array}\right\}\lb{sym2}
\end{eqnarray}
First, we chose $\varepsilon =2$. Thus, symmetry~$2)$ describes the
properties of the solution of Ding \& Hosoya \cite{DH}, equations
(\ref{RDH}), (\ref{wDH}).

{}From (\ref{rho1}) and (\ref{sym2}) we have
\begin{eqnarray}
R(\varrho ')&=&R(\varrho '')\lb{Rs1}\\
\dot R(\varrho ')&=&-\dot R(\varrho '')\lb{dRs1}\\
\ddot R(\varrho ')&=&\ddot R(\varrho '')\lb{ddRs1}\\
w(\varrho ')&=&-w(\varrho '')\lb{ws1}\\
\dot w(\varrho ')&=&\dot w(\varrho '')\lb{dws1}\\
\ddot w(\varrho ')&=&-\ddot w(\varrho '')\lb{ddws1}\\
V(\varrho ')&=&V(\varrho '')\lb{Vs1}\\
\frac{1}{2}\,V_{{}_{'}{w}}(\varrho ')&=&
-\frac{1}{2}\,V_{{}_{'}{w}}(\varrho '')\lb{dVs1}\\
\gamma (\varrho ')&=&-\gamma (\varrho '')\lb{gams1}\\
\dot\gamma (\varrho ')&=&\dot\gamma (\varrho '')\lb{dgams1}\:.
\end{eqnarray}
On the other hand, the conditions (\ref{Rs1}), (\ref{dRs1}),
(\ref{ws1}), (\ref{dws1}), (\ref{Vs1}), (\ref{dVs1}), (\ref{gams1}),
upon insertion into (\ref{w}) - (\ref{gamp}), must lead to the
equations (\ref{ddRs1}), (\ref{ddws1}), (\ref{dgams1}).
We will execute:
\begin{eqnarray*}
\ddot w(\varrho ')&=&\frac{1}{2}\,V_{{}_{'}{w}}(\varrho ')+
\frac12\,\dot w(\varrho ')\,\gamma (\varrho ')= \\
& &-\frac{1}{2}\,V_{{}_{'}{w}}(\varrho '')-
\frac12\,\dot w(\varrho '')\,\gamma (\varrho '')\,\,=\,\,-\ddot w(\varrho '')\\
\ddot R(\varrho ')&=&-\frac12\,\dot R(\varrho ')\,\gamma (\varrho ')-
G\,\frac{{\dot w}^2 (\varrho ')}{R(\varrho ')}= \\
& &-\frac12\,\dot R(\varrho '')\,\gamma (\varrho '')-
G\,\frac{{\dot w}^2 (\varrho '')}{R(\varrho '')}\,\,=\,\,\ddot R(\varrho '')\\
\dot\gamma (\varrho ')&=&\frac12\,\gamma ^2(\varrho ')-
\frac{2}{R^{\,2}(\varrho ')}\,\left[ \,2G\,V(\varrho ')+{\dot R}^{\,2}
(\varrho ')-1\,\right] = \\
& &\frac12\,\gamma ^2(\varrho '')-
\frac{2}{R^{\,2}(\varrho '')}\,\left[ \,2G\,V(\varrho '')+{\dot R}^{\,2}
(\varrho '')-1\,\right]\,\,=\,\,\dot\gamma (\varrho '')\:.
\end{eqnarray*}
Symmetry~1) is also easily explained. It corresponds to the
solution I have found numerically, as we will see in chapter~\ref{res}.

Now, we take an arbitrary $\varepsilon$. As before, we obtain:
\begin{eqnarray}
R(\varrho ')&=&R(\varrho '')\lb{Rs2}\\
\dot R(\varrho ')&=&-(\varepsilon -1)\dot R(\varrho '')\lb{dRs2}\\
\ddot R(\varrho ')&=&(\varepsilon -1)^2\ddot R(\varrho '')\lb{ddRs2}\\
w(\varrho ')&=&-w(\varrho '')\lb{ws2}\\
\dot w(\varrho ')&=&(\varepsilon -1)\dot w(\varrho '')\lb{dws2}\\
\ddot w(\varrho ')&=&-(\varepsilon -1)^2\ddot w(\varrho '')\lb{ddws2}\\
V(\varrho ')&=&V(\varrho '')\lb{Vs2}\\
\frac{1}{2}\,V_{{}_{'}{w}}(\varrho ')&=&
-\frac{1}{2}\,V_{{}_{'}{w}}(\varrho '')\lb{dVs2}\\
\gamma (\varrho ')&=&-\frac{1}{(\varepsilon -1)R(\varrho '')
\dot R(\varrho '')}\left[ \,2G\left( V(\varrho '')-(\varepsilon -1)^2
{\dot w}^2(\varrho '')\right) +\right. \nonumber\\
& & \left. \!+\left( (\varepsilon -1)^2{\dot R}^{\,2}(\varrho '')-1\right)
+\Lambda\,R^{\,2}(\varrho '')\,\right] \lb{gams2}\:.
\end{eqnarray}
If we insert (\ref{dws2}), (\ref{dVs2}), (\ref{gams2}) into
(\ref{w}), we see that an identity with (\ref{ddws2}) can be
achieved only if $\varepsilon =2$ applies. We don't need to
calculate any further. It won't help if (\ref{R}) and (\ref{gamp})
apply for an arbitrary $\varepsilon$, since (\ref{w}) doesn't.
And we have already observed the case $\varepsilon =2$.\\
{\bf {Summary:} }
\bd
\left.
\begin{array}{rcl}
R(\frac{\varrho}{2})=R(\varrho _{{}_0}-\frac{\varrho}{2})
& & R(\frac{\varrho}{2})=R(\varrho _{{}_0}-\frac{\varrho}{2})\\
w(\frac{\varrho}{2})=-w(\varrho _{{}_0}-\frac{\varrho}{2})
& & w(\frac{\varrho}{2})=w(\varrho _{{}_0}-\frac{\varrho}{2})
\end{array} \right\}\quad
\forall\varrho\in [0,\varrho _{{}_0}]\:.
\ed
Together with Prop.~\ref{prop5} and Prop.~\ref{prop6},
we are now able to sketch all possible solutions.
\newpage
%
\section{Numerics}
As we tried to solve the system (\ref{w}) - (\ref{Lp}) by means of a
standard routine for systems of ordinary dif\/ferential equations,
we concluded that most routines were unf\/it to do so. Only two
of them were satisfactory: DO2BAF from the NAG~library and
a routine taken from Ref.~\cite{SG}.

I tried to solve the equations (\ref{w}) - (\ref{dpot}) with
a standard shooting procedure for solving two-point boundary
value problems \cite{Press}, using both \cite{SG} and DO2BAF.
In the program taken from \cite{Press}, I just replaced the
routine for solving dif\/ferential equations, leaving the rest of the
program unchanged. The idea of the method mentioned in \cite{Press}
is to f\/ind such initial values (\ref{Re1}) - (\ref{we2}) that the
solutions meet at any point between the two boundary values.\\
Unfortunately, it turned out that the procedure was only
satisfactory enough when the initial values (\ref{Re1}) - (\ref{we2})
were close enough to the solution. At a distance of $10^{-2}$ from the
actual values, the computer was no longer able to f\/ind the solution.
I tried this with the already known solution (\ref{RDH}), (\ref{wDH}).

So I returned to the standard program. I constructed the initial
values with the expansions (\ref{Re1}) - (\ref{Le2}), at
$\varrho =10^{-5}$. By varying the parameters $b$ and $\Lambda$,
I shot for global solutions. The tolerance was of $10^{-8}$. The program
was supposed to run until $R=0$ or $|w|>1$. Now, I proceded as follows:
I f\/irst shot from the North Pole (i.e. at $\varrho =0$) and directed
$b$ and $\Lambda$ until the solution became regular. The initial
values came from the conditions (\ref{Re1}), (\ref{we1}), (\ref{ge1}),
(\ref{Le1}). The program told me now at which $\varrho$ the South
Pole was, i.e. how big $\varrho _{{}_0}$ was. Then I reversed the
routine and shot from the South Pole. The necessary initial values
now came from (\ref{Re2}), (\ref{we2}), (\ref{ge2}) and (\ref{Le2}).
$b$ and $\Lambda$ were virtually identical. It is not until we obtain
the same solutions by shooting from both sides that we can be sure
to have found a real solution.\\
At this point I also employed the routine taken from \cite{Press}, as the
values were close enough to the actual solution.
For this program, it was necessary to introduce a new independent
variable $t$, so that $\varrho$ became a function. I defined:
\begin{eqnarray*}
\varrho (t)&:=&\varrho\,t\;,\quad t\in [0,1]\:,\\
\varrho (0)&=&0\:,\\
\varrho (1)&=&\varrho _{{}_0}\:.
\end{eqnarray*}
Thus, now there are six parameters; namely $b$, $\Lambda$, $\varrho$
both for the North and the South Pole. Now I approximated very
accurately the parameters to the actual values.
\newpage
%
\section{Results}\lb{res}
As we have seen, the parameters of the solution of Ding \& Hosoya
\cite{DH} are
\begin{eqnarray*}
b&=&0.25\\
\Lambda&=&0.75\\
{\varrho}_{{}_0}&=&\sqrt{2}\,\pi\:,
\end{eqnarray*}
for $G=2$. The functions $R$, $w$, $\gamma$, $L$ are shown in
Fig.~\ref{figDH}.
\begin{figure}
\psfig{file=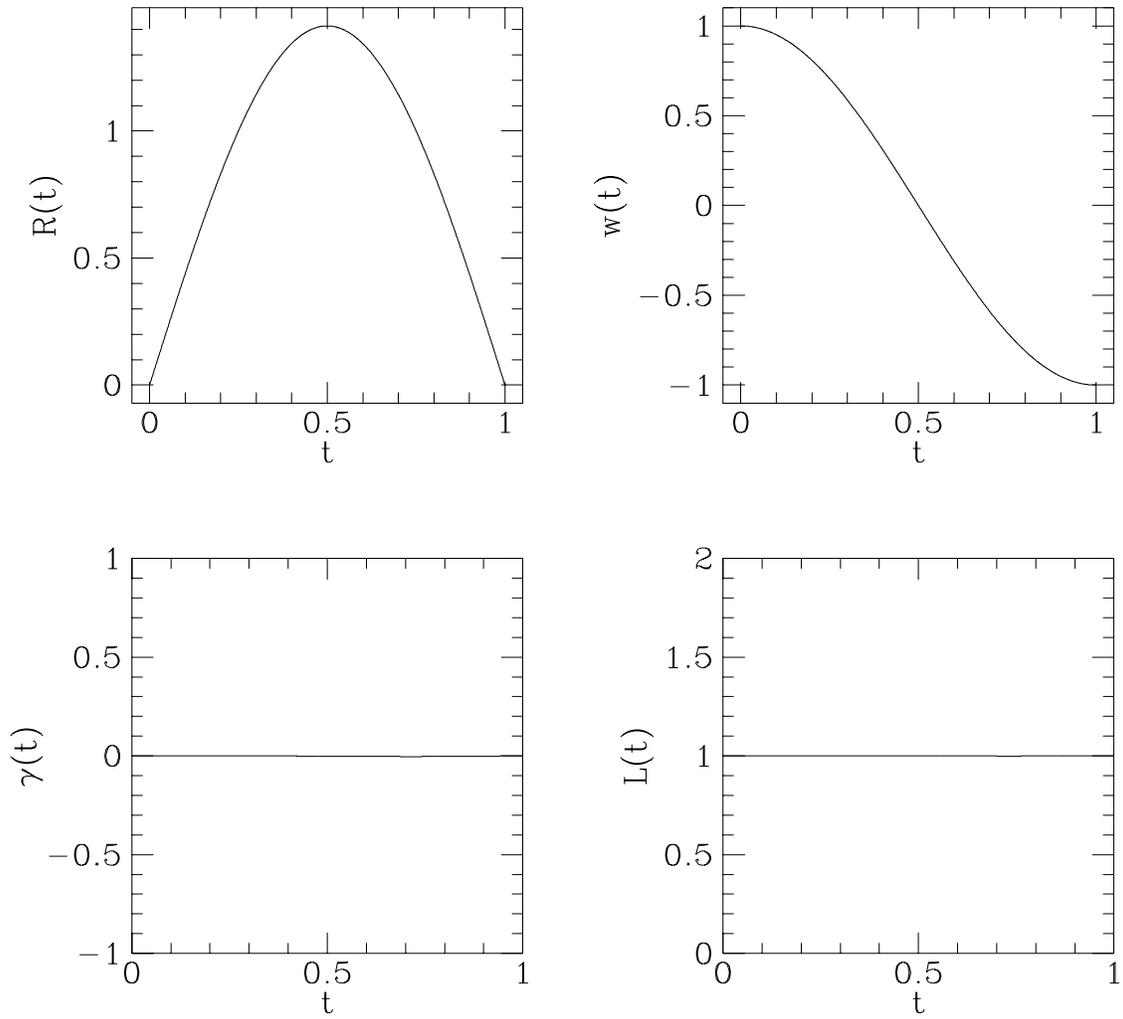,width=15cm,height=15cm}
\caption{The DH solution}\lb{figDH}
\end{figure}

If we assume $\Lambda =0.75$ and increase $b$, $w$ does not reach
the value $-1$; it rather moves towards zero again. An extremum
is formed. If we decrease $b$, $w$ becomes smaller than $-1$
and disappears, according to Prop.~\ref{prop1}, to inf\/inity.

If we assume a $\Lambda$ other than $0.75$ and try to arrange $b$
so as to allow a solution with a zero in $w$, we can observe the
following two cases:
\begin{description}
\item[$\Lambda >0.75$:]
In the best-case scenario, we can arrange $b$ such as to make
$R$ and $w$ look rather good. On the South Pole, however, there
is a peak that grows with an increasing $\Lambda$. In Fig.~\ref{figpeak},
you can recognize the peak. In the graph, $\dot R$ has been
drawn against $R$.
\item[$\Lambda <0.75$:]
In this case, there is a indentation at the South Pole. Again,
I have plotted $\dot R$ against $R$ (see Fig.~\ref{figind}).
\end{description}
\begin{figure}
\hbox to\hsize{
  \vbox{
     \psfig{file=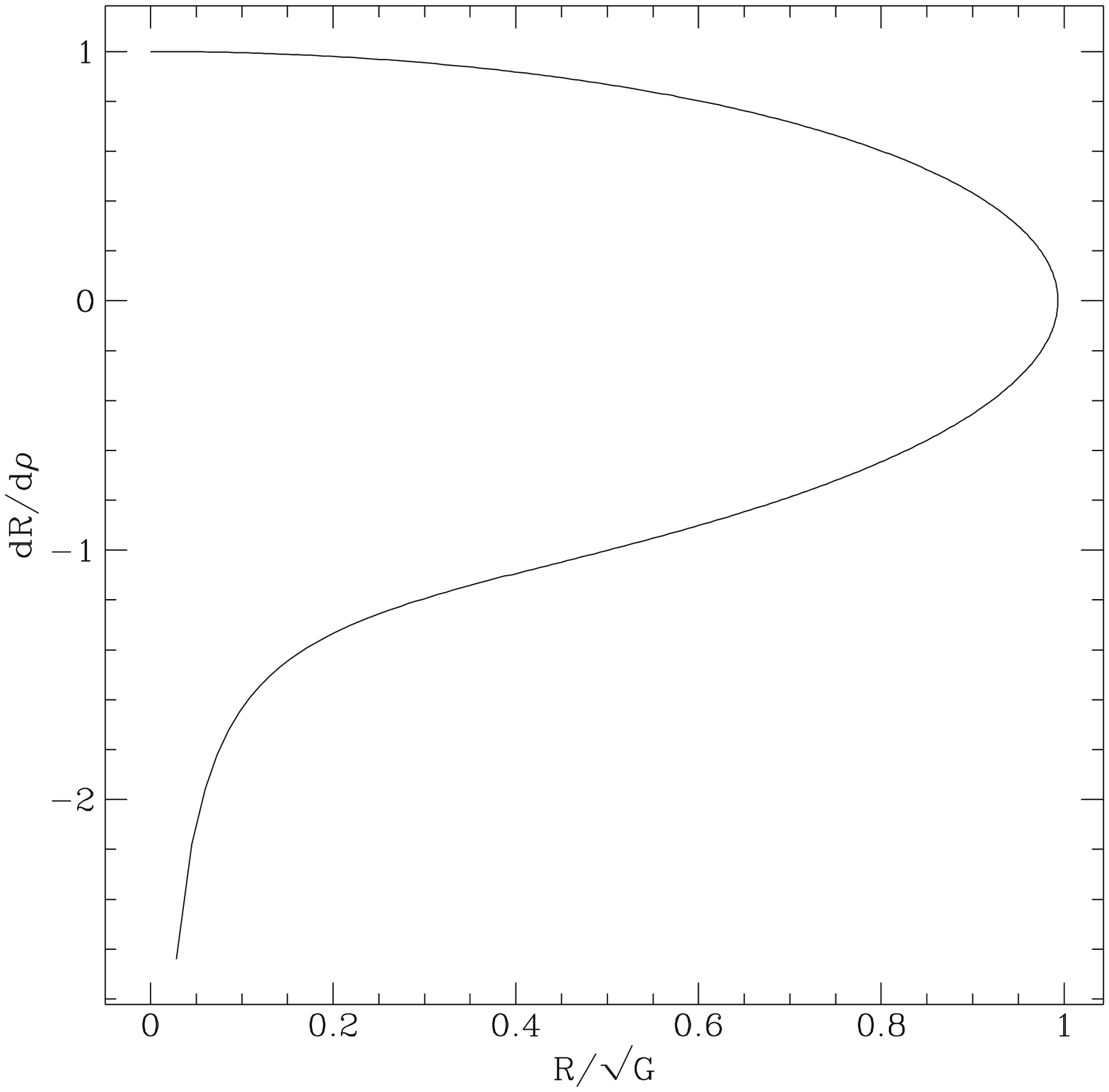,width=5cm,height=5cm}
     }
  \vbox{
     \psfig{file=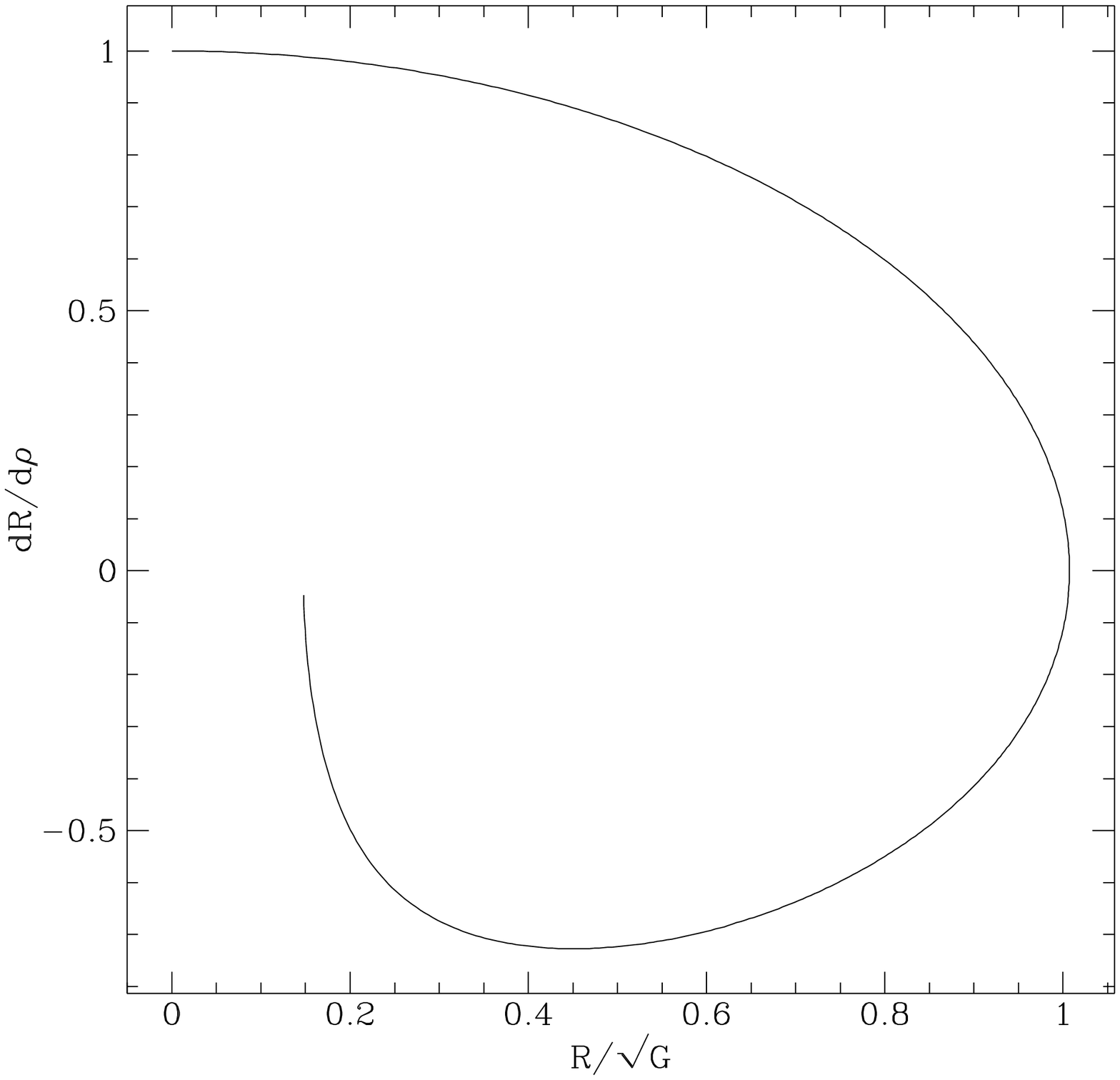,width=5cm,height=5cm}
     }}
\caption{$\dot R$ versus $R$ with $b=0.23655$, $\Lambda =0.8$}\lb{figpeak}
\caption{$\dot R$ versus $R$ with $b=0.2640625$, $\Lambda =0.7$}\lb{figind}
\end{figure}

In both cases the following applies: the farther away $\Lambda$
is from $0.75$ the farther the abscissa of the point with the
ordinate $-1$ moves of\/f from zero. Thus, I assert that for a zero
of $w$, there is only one $\Lambda$. In other words: for one zero,
there is only one regular solution. Now the question arises
whether there are also regular solutions with more than one zero
in $w$, and whether there is also just one regular solution for
any number of zeros. From the analogy with the asymptotically flat
case \cite{BK}, we can expect that.
And in fact, the same behavior was shown for two zeros, as
described above. I found a solution with the parameters
\begin{eqnarray*}
b&=&0.42976\nopagebreak[3] \\ \nopagebreak[3]
\Lambda&=&0.364\\ \nopagebreak[3]
{\varrho}_{{}_0}&=&8.64\:.
\end{eqnarray*}
The solution is shown in Fig.~\ref{figmy}. We see that the
symmetry~1) is fulf\/illed. There seem to be no regular solutions
with two zeros for other values of $\Lambda$.
\begin{figure}
\psfig{file=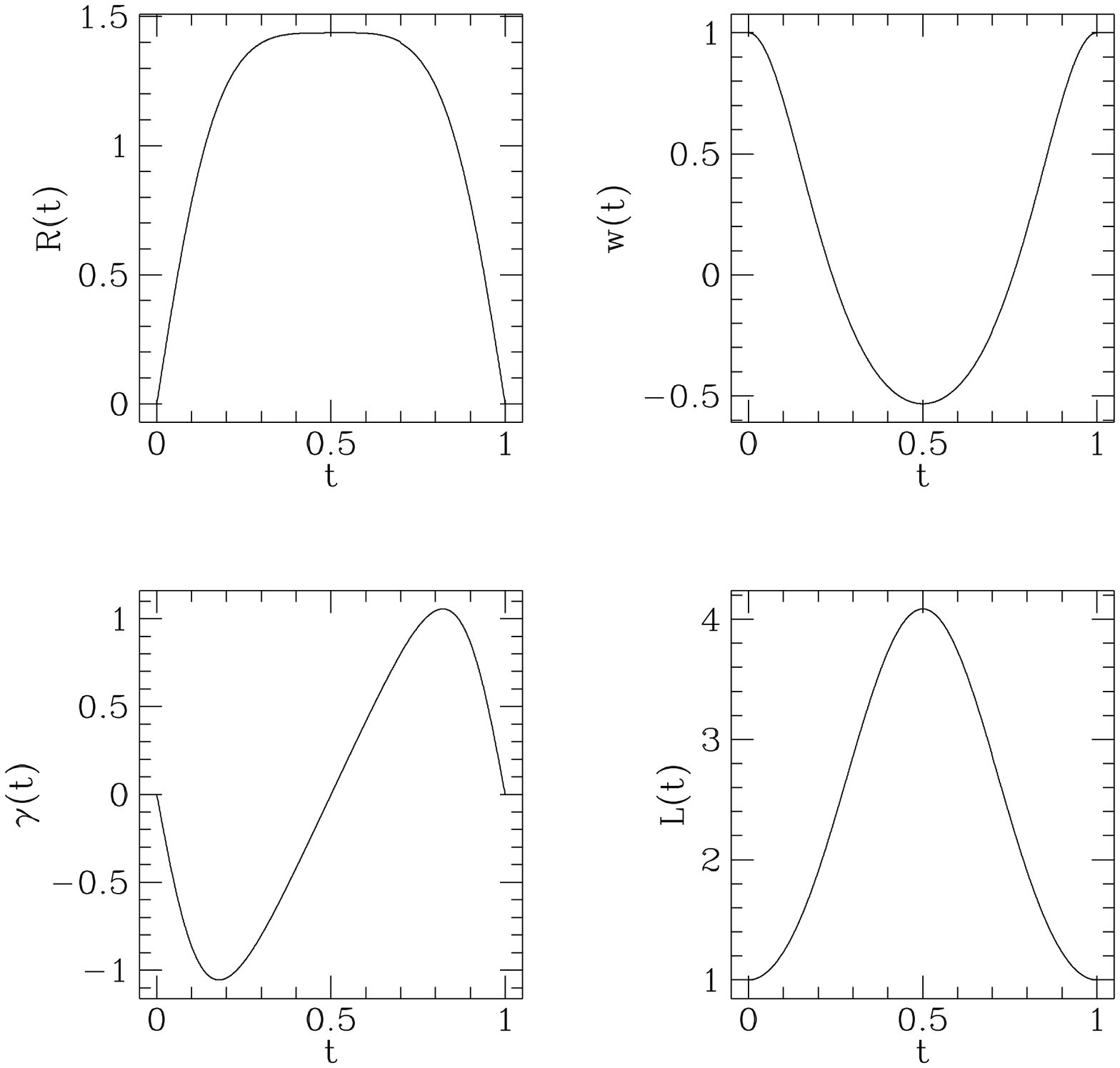,width=15cm,height=15cm}
\caption{The solution with $b=0.42976$, $\Lambda =0.364$,
$\varrho =8.64$}\lb{figmy}
\end{figure}

So far, I could not f\/ind any solutions with more than $2$ zeros,
either. We presume that either the numerics fail or a function in
the system becomes singular. As an example, we observe that the
functions
\begin{eqnarray*}
w&\equiv&1\:,\\
R&=&\sqrt{G}\,\sin \left( \frac{\varrho}{\sqrt{G}}\right) \:,\\
\gamma &=& \frac{2}{\sqrt{G}}\,\tan \left( \frac{\varrho}{\sqrt{G}}\right)
\:,\\
L&=&\cos \left( \frac{\varrho}{\sqrt{G}}\right)
\end{eqnarray*}
solve our system (\ref{w}) - (\ref{Lp}). $\gamma$
becomes, however, singular at the equator. Thus, in our coordinates we don't
f\/ind the solution on the computer, although it solves our system.
That could mean that we cannot f\/ind all solutions in these
coordinates. We must try to introduce more appropriate coordinates.
\newpage
%
\section{Time dependent cosmological solutions to the EYM system}
Our metric shall be
\be\lb{met5}
g=a^2\!(t)\!\left[ -dt^2\!+h\right] \:,
\ee
with $h$ being the standard metric on $S^3$. The radius will be taken
equal to $1$, which is always possible. We choose the gauge potential
\be\lb{eichpot5}
A=f\theta \:,
\ee
where $\theta$ is the Maurer-Cartan form on $S^3$, understood as SU$(2)$.
$f$ shall depend exclusively upon time: $f=f(t)$.

We set up the coupled f\/ield equations:\\
For the YM equations we can use the metric $\hat g=\left( -dt^2+h\right)$,
since they are conformally invariant. It is helpful to express the metric $h$
in terms of $\theta$. Let
\begin{eqnarray}
\theta & = & \sum_{i=1}^{3} \theta ^i \tau _i \qquad (\vec \tau =
\vec \sigma /i) \lb{thetar5}\\
\theta ^0 & = & dt \lb{theta05}\:;\\
\mbox{then}\quad h&=&\sum \left( \theta ^i \right) ^2\:,
\quad \hat g =\eta_{\mu\nu}\,\theta ^{\mu}\!\otimes \theta ^{\nu}
\lb{meth5}\:.
\end{eqnarray}
With the help of the Maurer-Cartan equations
\be\lb{mc5}
d\theta\,+\,\theta \wedge \theta =d\theta\,+\,\frac12 [\theta ,\theta ]=0
\ee
we f\/ind the f\/ield strenght tensor
\be\lb{F5}
F=\dot f \,\theta ^0 \!\wedge \theta  \,+
\,(f^2\!-f)\,\frac12 [\theta ,\theta ]
\ee
or
\begin{eqnarray}
F & = & \dot f \,\theta ^0 \!\wedge \theta ^1 \tau _1 \,+\,
\dot f \,\theta ^0 \!\wedge \theta ^2 \tau _2 \,+\,
\dot f \,\theta ^0 \!\wedge \theta ^3 \tau _3
\nonumber\\
& & +2(f^2 \!-f)\left[ \theta ^2 \!\wedge \theta ^3 \tau _1 \,+\,
\theta ^3 \!\wedge \theta ^1 \tau _2 \,+\,
\theta ^1 \!\wedge \theta ^2 \tau _3 \right]\lb{Fa5}\:.
\end{eqnarray}
The upper line of (\ref{Fa5}) conteins the electrical components,
and the lower one the magnetic components. Now, (\ref{Fa5})
gives us very quickly
\be \lb{skp5}
(F|F)=\left[ -{\dot f}^{\,2}\,+\,4\left( f^2 \!-f\right) ^2\right] \cdot 3
\quad .
\ee
Here we have used the following normalization of the
scalar product for the Lie~algebra
\bd
\langle X,Y\rangle =-\frac12\,tr(X\!\cdot Y)\qquad (\mbox{i.e.}\,
\langle \tau _{i},\tau _{j}\rangle =\delta _{ij})\:.
\ed
Thus, the YM equations are reduced to a 1-di\-men\-sio\-nal
``mechanical pro\-blem'' with the Lagrange function
\be \lb{lag5}
L_{{}_{YM}}={\dot f}^{\,2}-\,4\left( f^2\!-f\right) ^2\equiv T-V\:.
\ee
The corresponding energy is $T+V=E$:
\be \lb{energ5}
{\dot f}^{\,2}+\,4\left( f^2\!-f\right) ^2 =E\:.
\ee
In Fig.~\ref{figv}, we plot the potential $V$ and also the phase portrait
(Fig.~\ref{figpt}).
There are two areas of stability at $f=0$ and $f=1$, while exactly
in the middle, at $f=1/2$, we have an unstable point, which
turn out to be gauge equivalent to the DH solution as already mentioned
in the introduction. Now, we explain this in short.
\begin{figure}
\hbox to\hsize{
  \vbox{
     \psfig{file=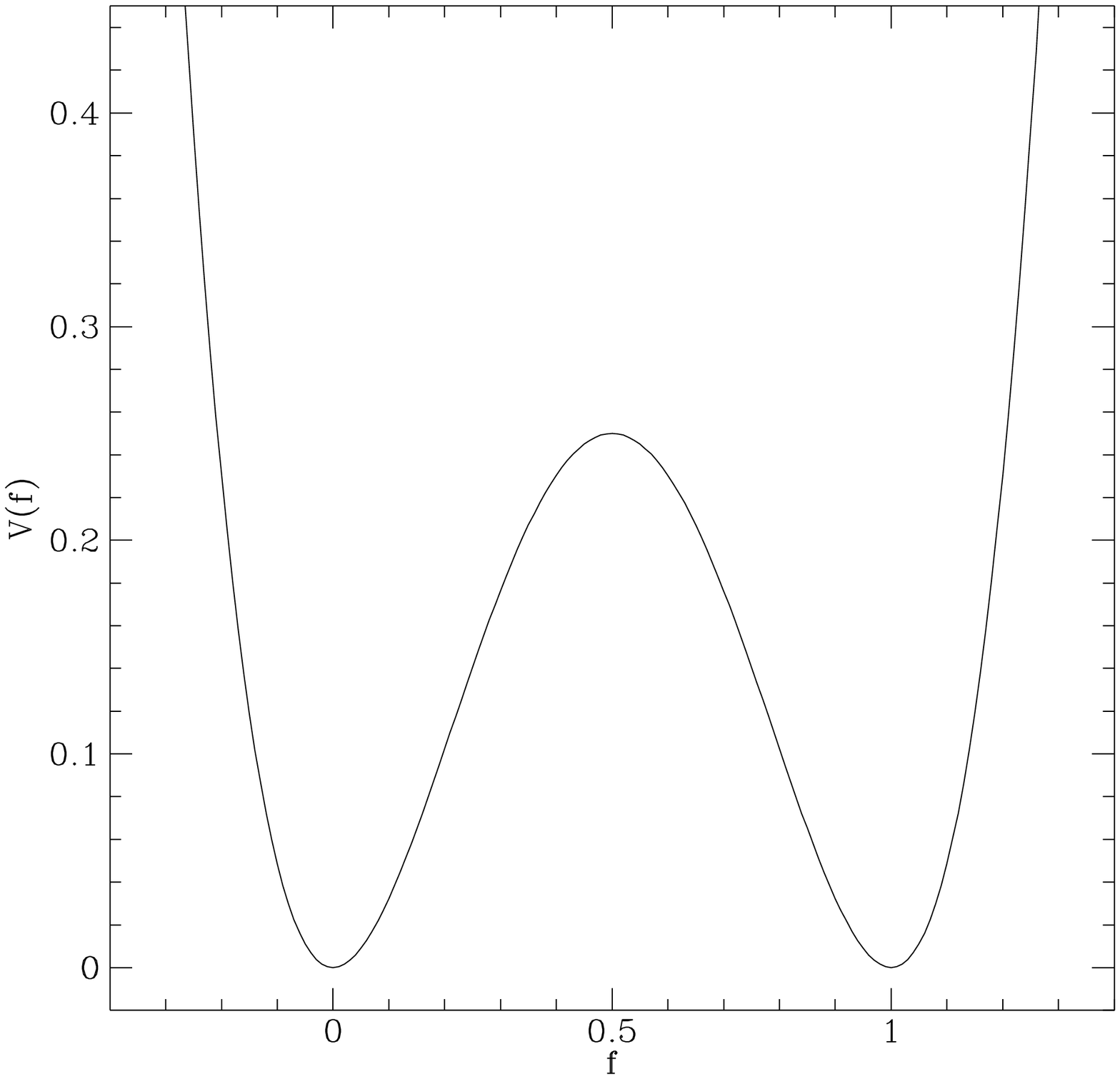,width=5cm,height=5cm}
     }
  \vbox{
     \psfig{file=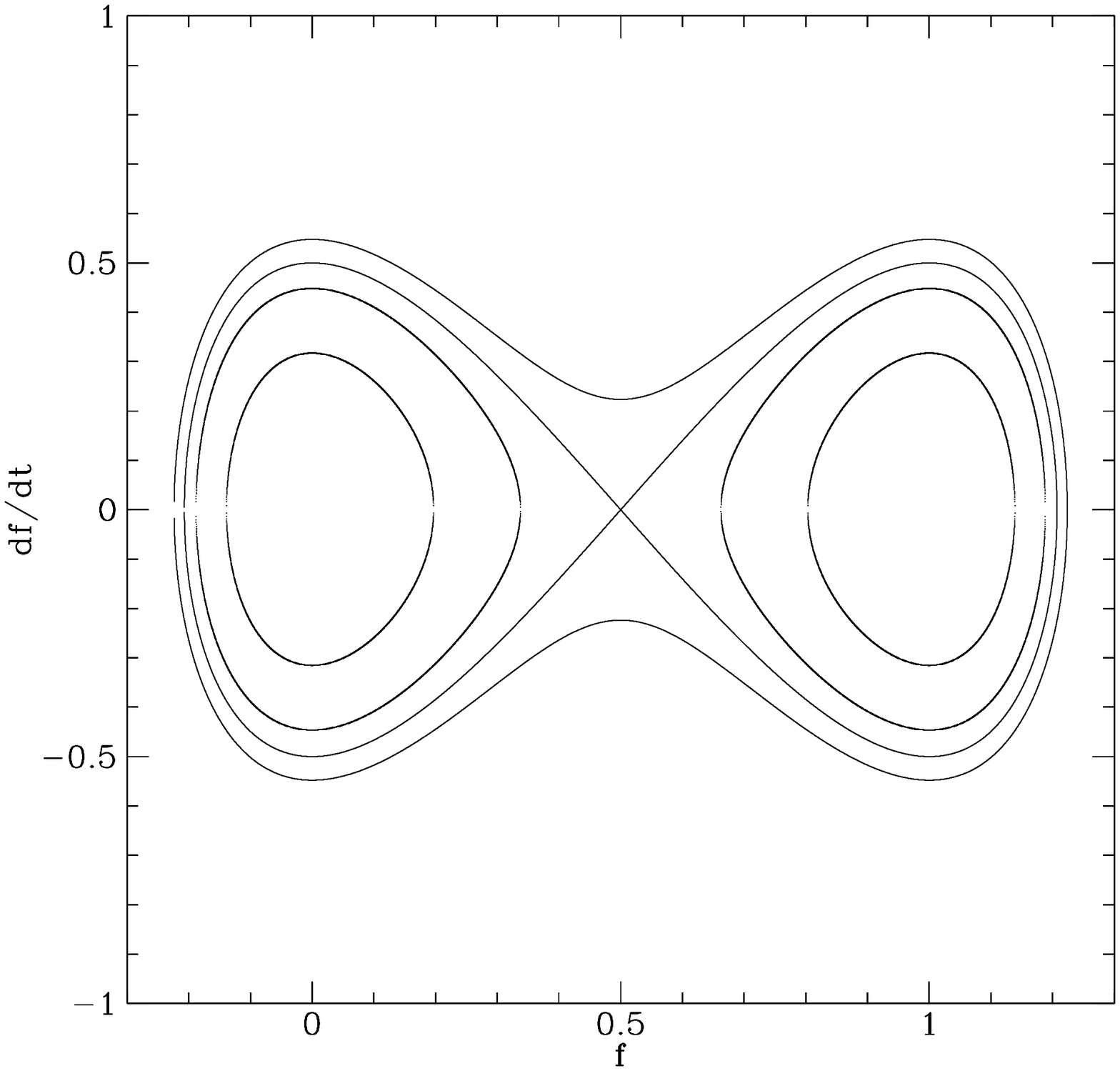,width=5cm,height=5cm}
     }}
\caption{The potential $V(f)$}\lb{figv}
\caption{The phase portrait with $E=0.1$, $0.2$, $0.25$, $0.3$}\lb{figpt}
\end{figure}
We def\/ine the following map
\bd
g:S^{\,3}\to SU(2)\:,\quad x \mapsto x^4 \cdot 1+i\,\vec x \,\vec{\sigma}\:,
\ed
\bd
x \in {\rm R}^4\:, \quad |x|=1\:,
\ed
and the one-form
\bd
\theta =g^{-1}dg\:.
\ed
We choose the adequate coordinates
\bd
r=|x|\:, \quad x^4 =\sqrt{1-r^2}\:.
\ed
Then the following applies
\bd
\theta =\left( x^4 \cdot 1-i\,\vec{\sigma}\,\vec x \right)
\left( dx^4 +i\,\vec{\sigma}\,d\vec x \right) \:.
\ed
After a short calculation we obtain
\bd
\frac{1}{i}\,\theta =\vec{\sigma}\,\left\{ \sqrt{1-r^2}\,d\vec x
+\vec x \,\frac{\vec x \,d\vec x }{\sqrt{1-r^2}}+\vec x \wedge d\vec x \right\}
\ed
and f\/inally $(\vec{\tau}=\vec{\sigma}/i)$
\begin{eqnarray*}
-\theta &=& \frac{1}{\sqrt{1-r^2}}\,\tau _{r}\,dr+r\,\sqrt{1-r^2}\,
\Bigl( \tau _{\vartheta}\,d\vartheta +\tau _{\varphi}\,\sin\!\vartheta\,
d\varphi \Bigr) \\
& & -r^2\,\Bigl( \tau _{\varphi}\,d\vartheta -\tau _{\vartheta}\,
\sin\!\vartheta \,d\varphi \Bigr) \:.
\end{eqnarray*}
If we def\/ine $r=\sin\!\chi $, it follows that
\begin{eqnarray*}
-\theta &=& \tau _{r}\,d\chi +\sin\!\chi\,\cos\!\chi\,
\Bigl( \tau _{\vartheta}\,d\vartheta +\tau _{\varphi}\,\sin\!\vartheta\,
d\varphi \Bigr) \\
& & -\sin ^2\!\chi\,\Bigl( \tau _{\varphi}\,d\vartheta -\tau _{\vartheta}\,
\sin\!\vartheta \,d\varphi \Bigr) \:.
\end{eqnarray*}
The BK ansatz is
\bd
\tilde A =\frac{w-1}{2}\,\Bigl[ \tau _{\varphi}\,d\vartheta -\tau
_{\vartheta}\,
\sin\!\vartheta \,d\varphi \Bigr] \:.
\ed
We transform this with $\,G=\cos\!\tilde\chi +\tau _r\sin\!\tilde\chi \,$
and obtain $(\lambda\equiv 2\tilde\chi )$:
\begin{eqnarray*}
G^{-1}A\,G+G^{-1}\,dG &=& \frac12\,\tau _r\,d\lambda +
\frac{w\,\cos\!\lambda -1}{2}\,\Bigl( \tau _{\varphi}\,d\vartheta -\tau
_{\vartheta}\,\sin\!\vartheta \,d\varphi \Bigr) \\
& & +\frac{w}{2}\,\sin\!\lambda\,\Bigl( \tau _{\vartheta}\,d\vartheta +\tau
_{\varphi}\,\sin\!\vartheta\,d\varphi \Bigr) \:.
\end{eqnarray*}
This is equal to $\,\frac12 \,\theta \,$ for $\,w=\cos\!\lambda \,$ and $\,\chi
=\lambda \,$.
Thus, we f\/ind indeed
\bd
\tilde A=\frac12 \,\theta \:.
\ed

We set up the Einstein equations with $\Lambda$. Since
$tr(T)=0$, we consider f\/irst the trace of the
Einstein equations: $R=4\Lambda$. Since
\bd
R(g)=\frac{6}{a^2}\left[ \frac{\ddot a}{a}+1\right]\:,
\ed
we obtain
\be \lb{tr5}
\ddot a +a=\frac{2\Lambda}{3}\,a^3\:.
\ee
With respect to the orthonormalized tetrade $\{a\theta\}$ of $g$,
$G_{00}=\frac{3}{a^2}\,\left[ \left( \frac{\dot a}{a}\right) ^2
+1\right]$. The energy density is given by (use (\ref{Fa5})):
\be \lb{energd5}
8\pi T_{00}=\left[ \frac{{\dot f}^{\,2}}{a^4}\cdot 3
+\frac{4\left( f^2\!-f\right) ^2}{a^4}\cdot 3\right] =\frac{3}{a^4}(T+V)\:.
\ee
Thus, the Friedman equation with the $\Lambda$-term is:
\be \lb{Fried5}
{\dot a}^{2}\!+a^2 =T+V+\frac{\Lambda}{3}\,a^4\:.
\ee
For every $T+V=E$, this is again a ``mechanical problem'' for $a(t)$:
\be \lb{mech5}
{\dot a}^{2}\!+U(a)=E\;,\quad U(a)=a^2\!-\frac{\Lambda}{3}\,a^4\:.
\ee
Let us f\/irst consider the case $\Lambda =0$. Then
\bd
\ddot a +a=0\;,\quad {\dot a}^{2}\!+a^2 =E\:,
\ed
hence
\be \lb{at5}
a(t)=a_{{}_0}\sin t\:.
\ee
When we take $f$ as being static, the derivative of (\ref{energ5})
will lead us to:
\bd
f(f\!-1)(2f\!-1)=0\:.
\ed
We obtain three solutions:
\be \lb{3sol5}
f_1 =0\;,\quad f_2 =1\;,\quad f_3 =\frac12 \;.
\ee
For $f_1$, $A$ equals zero, and $f_2$ corresponds to a pure gauge
$A=\theta$. For the interesting case $f=1/2$, $E$ equals $1/4$.
If we shift to the physical time $\tilde t$, $\left( d\tilde t =a\,dt\right) $,
this leads to
\be \lb{time5}
a(\tilde t )={\tilde t}^{{}^{\,1/2}}\left( 2a_{{}_0}-\tilde t\right) ^{1/2}\:.
\ee
There is a static solution for $\Lambda\ne 0$, namely $f=1/2$,
$a=(2\,\Lambda /3)^{-1/2}$ (this follows from (\ref{tr5})).
In Fig.~\ref{figu}, we see the shape of $U(a)$. The $a$ above is
exactly the critical point of $U(a)$. The DH solution corresponds
to the local maximum of the potential $V(f)$ and is therefore
{\em{\/unstable\/}}.
\begin{figure}
\hbox to\hsize{
  \psfig{file=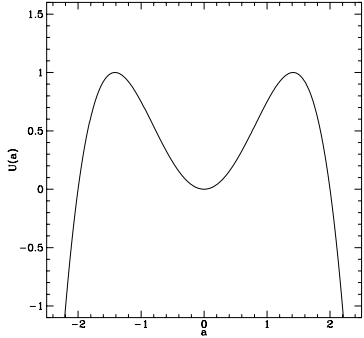,width=5cm,height=5cm}
  }
\caption{The potential $U(a)$ with $\Lambda =3/4$}\lb{figu}
\end{figure}

A stability analysis of the new solution, as well as other material
will be published elsewhere \cite{HLMS}.
\section*{Acknowledgments}
During the work for my diploma thesis, some doubts arose about
whether positive results would come out. In order to complete my
work, I was given a further assignment. I thank Prof.~Dr.~N.~Straumann
for this support. Discussions with Othmar
Brodbeck, Mikhail Volkov and George Lavrelashvili also
proved to be very helpful. I am also very obliged to Markus
Heusler. In the beginning, he introduced me into the matter
of subject with great patience and he has helped me every
time I had specif\/ic questions. F\/inally, my thanks go to
Ivan Colaci for his translation of my German draft into English.

%


\begin{thebibliography}{99}

\newcommand{\PL}{{\em \/Phys. Lett.\/} }
\newcommand{\PRL}{{\em \/Phys. Rev. Lett.\/} }
\newcommand{\JMP}{{\em \/J. Math. Phys.\/} }
\newcommand{\CMP}{{\em \/Commun. Math. Phys.\/} }
\newcommand{\NP}{{\em \/Nucl. Phys.\/} }


\bibitem{Li} A. Lichnerowicz, in {\em\/Th\'{e}ories relativistes de la
gravitation et de l'\'{e}l\'{e}ctro\-magne\-tisme\/} (Masson, Paris, 1955)

\bibitem{De} S. Deser, \PL {\bf B 64}, 463 (1976)

\bibitem{SC} S. Coleman, in {\em\/New Phenomenon in Subnuclear Physics\/},
ed. A. Zichichi (Plenum, New York, 1975)

\bibitem{BK} R. Bartnik and J. McKinnon, \PRL {\bf 61}, 141 (1988)

\bibitem{VG} M. S. Volkov and D. V. Galt'sov, {\em Prs'ma Zh. Eksp.
Teor. Fiz.} {\bf 50}, 312 (1989);  {\em Sov. J. Nucl. Phys.} {\bf 51},
747 (1990)

\bibitem{PB} P. Bizon, \PRL {\bf 64}, 2844 (1990)

\bibitem{KM} H. P. K\"unzle and A. K. Masoud-ul-Alan, \JMP {\bf
31}, 928 (1990)

\bibitem{SWY1} J. A. Smoller, A. G. Wassermann, S. T. Yau and J. B. McLeod,
\CMP {\bf 143}, 115 (1992)

\bibitem{SW} J. A. Smoller and A. G. Wassermann, \CMP {\bf 151}, 303 (1993)

\bibitem{SWY2} J. A. Smoller, A. G. Wassermann and S. T. Yau, \CMP
{\bf 154}, 377 (1993)

\bibitem{Breit} P. Breitenlohner, P. Forg\'{a}cs and D. Maison, \CMP
{\bf 163}, 141 (1994)

\bibitem{SZ1} N. Straumann and Z.-H. Zhou, \PL {\bf B 237}, 353 (1990)

\bibitem{SZ2} N. Straumann and Z.-H. Zhou, \PL {\bf B 243}, 33 (1990)

\bibitem{SZ3} Z.-H. Zhou and N. Straumann, \NP {\bf B 360}, 180
(1991)

\bibitem{Z} Z.-H. Zhou, {\em Helv. Phys. Acta} {\bf 65}, 767 (1992)

\bibitem{OS1} O. Brodbeck and N. Straumann, \PL {\bf B 324}, 309 (1994)

\bibitem{OS2} O. Brodbeck and N. Straumann, {\em\/Instability Proof
for Einstein-Yang-Mills Solitons and Black Holes with arbitrary
Gauge Groups\/}, Z\"urich University Preprint No. ZU-TH 38/1994,
gr-qc/9411058

\bibitem{DHS1} S. Droz, M. Heusler and N. Straumann, \PL {\bf B 268},
371 (1991)

\bibitem{DHS2} M. Heusler, S. Droz and N. Straumann, \PL {\bf B 271},
61 (1991)

\bibitem{DHS3} M. Heusler, S. Droz and N. Straumann, \PL {\bf B 285},
21 (1992)

\bibitem{HS} M. Heusler, N. Straumann and Z.-H. Zhou, {\em Helv. Phys. Acta}
{\bf 66}, 614 (1993)

\bibitem{LNW} K.-Y. Lee, V. P. Nair and E. Weinberg, \PRL {\bf 68}, 1100
(1992); {\em\/Phys. Rev.\/} {\bf D 45}, 2751 (1992); M. E. Ortiz,
{\em\/ibid.\/} {\bf 45}, R2586 (1992)

\bibitem{BFM} P. Breitenlohner, P. Forg\'{a}cs and D. Maison, \NP {\bf 383},
357 (1992)

\bibitem{AB} P. C. Aichelburg and P. Bizon, {\em\/Phys. Rev.\/} {\bf D 48},
607 (1993)

\bibitem{DG} E. E. Donets and D. V. Gal'tsov, \PL {\bf B 302}, 411 (1993);
{\bf 312}, 391 (1993); G. Lavrelashvili and D. Maison, \NP {\bf B 410},
407 (1993); C. M. O'Neill, Institution Report No. CLNS-93/1246 and
\mbox{hep-th/9311022}, 1993 (unpublished); P. Bizon, {\em\/Act. Phys. Pol.\/}
{\bf B 24}, 1209 (1993)

\bibitem{BBMSV} P. Boschung, O. Brodbeck, F. Moser, N. Straumann and
M. S. Volkov, {\em\/Phys. Rev.\/} {\bf D 50}, 3842 (1994)

\bibitem{GMN} B. R. Greene, S. D. Mathur and C. M. O'Neill,
{\em\/Phys. Rev.\/} {\bf D 47}, 2242 (1993)

\bibitem{DH} S. Ding and A. Hosoya, TIT/HEP-242/COSMO-39, 1993

\bibitem{com} private communication by N.~Straumann

\bibitem{YH} Y. Hosotani, \PL {\bf B 147}, 44 (1984)


\bibitem{NS} N. Straumann, {\em\/Allgemeine Relativit\"atstheorie
und relativistische Astrophysik\/}, Springer-Verlag (1988), Vol. 150, p. 441

\bibitem{St} H. St\"ocker, {\em\/Taschenbuch mathematischer Formeln
und moderner Verfahren\/}, Verlag Harri Deutsch (1993), p. 113

\bibitem{SG} L. F. Shampine, M. K. Gordon, {\em\/Computer Solutions
of Ordinary Differential Equations: The Initial Value Problem\/},
W. H. Freeman and company, San~Francisco, 1975

\bibitem{Press} W. H. Press et al., {\em\/Numerical Recipies\/},
Cambridge University Press, New~York, 1992

\bibitem{HLMS} M. Heusler, G. Lavrelashvili, P. Moln\'{a}r and
N. Straumann, in preparation

\end{thebibliography}
\end{document}